\documentstyle[epsfig,natbib2,natbibmnfix]{mn2e}

\bibpunct{(}{)}{;}{a}{}{,}

\def\msun{{\,{\rm M}_\odot}}

\def\sgra{Sgr~A$^*$}

\def\simlt{\lower.5ex\hbox{$\; \buildrel < \over \sim \;$}}
\def\simgt{\lower.5ex\hbox{$\; \buildrel > \over \sim \;$}}
\def\del#1{{}}
\newcommand\schw{Schwarzschild}

\renewcommand{\vec}[1]{ {\bmath #1} } 

\title{Galactic Centre stellar winds and Sgr A* accretion}

\author[J.~Cuadra, S.~Nayakshin, V.~Springel, \& T.~Di~Matteo]
{\parbox{18cm}{Jorge Cuadra\footnotemark[3], Sergei Nayakshin\footnotemark[4],
Volker Springel \& Tiziana Di Matteo\footnotemark[2]}\vspace{0.3cm}\\
Max-Planck-Institut f\"{u}r Astrophysik, Karl-Schwarzschild-Stra\ss{}e 1,
85741 Garching bei M\"{u}nchen, Germany}

\begin{document}
\maketitle

\begin{abstract}
We present a detailed discussion of our new 3D numerical models for
the accretion of stellar winds on to \sgra.  In our most sophisticated
models, we put stellar wind sources on realistic orbits around \sgra,
we include recently discovered `slow' winds ($v_{\rm w}\sim 300$ km
s$^{-1}$), and we account for optically thin radiative cooling. We
test our approach by first modelling only one phase `fast' stellar
winds ($v_{\rm w}\sim 1000$ km s$^{-1}$).  For stellar wind sources
fixed in space, the accretion rate is of the order of $\dot M \simeq
10^{-5} \msun$~yr$^{-1}$, fluctuates by $\simlt 10$\%, and is in a
good agreement with previous models.  In contrast, $\dot M$ decreases
by an order of magnitude for wind sources following circular orbits,
and fluctuates by $\sim 50$\%. Then we allow
 a fraction of stars to produce slow winds. Much of
these winds cool radiatively after being shocked, forming cold clumps
and filaments immersed into the X-ray emitting gas.  We investigate
two orbital configurations for the stars in this scenario, an
isotropic distribution and two rotating discs with perpendicular
orientation. The morphology of cold gas is quite sensitive to the
orbital distribution of the stars.  In both cases, however, most of
the accreted gas is hot, producing a quasi steady `floor' in the
accretion rate, of the order of $\sim 3\times 10^{-6}
\msun$~yr$^{-1}$, consistent with the values deduced from {\em
Chandra} observations.  The cold gas accretes in intermittent, short
but powerful accretion episodes which may give rise to large amplitude
variability in the luminosity of \sgra\ on time scales of tens to
hundreds of years.  The circularisation radii for the flows are about
$10^3$ and $10^4$ Schwarzschild radii, for the one and two-phase wind
simulations, respectively, never forming the quasi-spherical accretion
flows suggested in some previous work. Our work suggests that,
averaged over time scales of hundreds to thousands of years, the
radiative and mechanical luminosity of \sgra\ may be { 
substantially} higher than it is in its current state. Further improvements
of the wind accretion modelling of \sgra\ will rely on improved
observational constraints for the wind velocities, mass loss rates and
stellar orbits.
\end{abstract}

\renewcommand{\thefootnote}{\fnsymbol{footnote}}
\footnotetext[3]{E-mail: {\tt jcuadra@mpa-garching.mpg.de}}
\footnotetext[4]{Current address: Department of Physics \& Astronomy,
University of Leicester, Leicester LE1 7RH, UK} 
\footnotetext[2]{Current
address: Department of Physics, Carnegie Mellon University,
Pittsburgh, PA 15213, USA}

\section{Introduction}

\sgra\ is believed to be a super-massive black hole (SMBH) of mass
$M_{\rm BH} \simeq 3.5 \times 10^6 \msun$ in the very centre of our
Galaxy \citep[e.g., ][]{Reid99,Schoedel02,Ghez03b}. Winds from young
massive stars with velocity $v_{\rm w}$ around 1000 km s$^{-1}$ are
known to fill the inner parsec \citep[e.g.,][]{Hall82} with hot
plasma. The total mass loss rate from these stars is $\sim 10^{-3}
\msun$ yr$^{-1}$ \citep[e.g.,][]{Genzel94, Najarro97}, and a fraction
of this gas should be accreting on to the SMBH
\citep[e.g.,][]{Melia92}. The observed luminosity is however many
orders of magnitude smaller than what is predicted from the classical
Bondi-Hoyle theory \citep{Bondi52}. There are two possible
explanations for this discrepancy: either a much smaller amount of gas
actually accretes on to the SMBH or accretion proceeds in a
low-radiative efficiency mode. The current consensus appears to be
that both of these factors are important for reducing \sgra's
luminosity \citep{Narayan02}. From theoretical arguments, it is
unlikely that the accretion flow is exactly spherical, and instead it
is plausible that a rotating flow forms in which the resulting viscous
or convective heating unbinds much of the gas, severely reducing the
accretion rate \citep{Blandford99, Quataert00}.  It is also likely
that electrons are not as hot as the ions, thus resulting in a greatly
diminished radiative efficiency of the flow \citep{NY94}.

Observations of \sgra\ in the radio and in the X-ray bands constrain
the accretion rate at tens of \schw\ radii distance from the SMBH to
values of the order of $\dot M_{\rm in}\sim$~few$\times 10^{-7}
\msun$~year$^{-1}$ \citep{Bower03, Nayakshin05}. This is significantly
smaller than the $\sim 3\times 10^{-6} \msun$~year$^{-1}$ accretion
rate estimated based on {\em Chandra} observations at $\sim 10^5$
\schw\ radii \citep{Baganoff03a}, confirming that gas outflows are
important.  For the sake of closer testing accretion flow theories, it
is important to establish the exact amount of gas captured by \sgra\
to compare to $\dot M_{\rm in}$. Note that this exercise can be done
only for \sgra\ at present, since all other AGN are much farther away,
and hence \sgra\ is a unique test object in this regard.

The first three dimensional numerical simulations of \sgra\  wind
accretion were performed {  by \cite{Ruffert94}, who studied feeding the black
hole from a uniform large scale gas flow. This work was extended by \cite{Coker97},
who used discrete gas sources, ten mass-loosing stars semi-randomly
positioned a few arc-seconds away from \sgra, to model the wind
emission.} Due to numerical difficulties inherent to fixed grid codes,
the orbital motions of the stars could not be followed, and thus they
were fixed in space. The authors argued that such an approach is valid
since the wind velocities, as best known at the time, are larger than
the circular Keplerian motions of the stars in these
locations. Moreover, if the stellar orbits are isotropically
distributed, and all stars are identical, then the net angular
momentum is nearly zero. More recently, \cite{Rockefeller04} used a
particle-based code, making also use of more detailed information on
stellar coordinates and wind properties. However, the stars once again
were kept at fixed locations.

Recent near-IR data of the nuclear star cluster in \sgra\ show that
the stellar wind sources are located in two ring- or disc-like
distributions that are roughly perpendicular to each other
\citep{Paumard01, Genzel03a}, implying that shocked gas has a non-zero
net angular momentum. This is likely to be important for understanding
the structure of the accretion flow. In addition, integral field
spectroscopy of the central parsec implies stellar wind velocities of
only $\sim 200$ km s$^{-1}$ \citep{Paumard01}; much less than the
values of $\sim 600$ km s$^{-1}$ previously estimated. These new data
imply that stellar orbital motions are more important than previously
thought. Moreover, these slow winds, when shocked, are heated to
around $10^6$ K as opposed to the temperature of $1-2 \times 10^7$ K
for the hotter $\sim 1000$ km s$^{-1}$ winds. This former slow phase
of \sgra\ stellar winds is therefore susceptible to radiative cooling
\citep{C05} and thus it is expected to form a {\em cold} gas flow on to
\sgra\ in addition to the usually studied hot non-radiative phase.

Motivated by the new observations and the above ideas, we performed
numerical simulations of wind accretion on to \sgra\ including
optically thin radiative cooling and allowing the wind-producing stars
to be on circular Keplerian orbits.  Some preliminary results of our
study have already been presented by \cite{C05}. Here we report on
specific tests of our new methodology and provide further details on
our results. While this study sheds new light on physics of accretion
of stellar winds on to \sgra, it is clear that improved observational
determinations of stellar mass loss rates, wind velocities, stellar
orbits, and also of orbits and distribution of the cooler gas phase
filling in the inner parsec will be the key for further improving our
understanding of the Galactic centre region.

The paper is structured as following.  We describe our numerical
method in Section~\ref{sec:method}, and give results of simulations
with single, fast wind velocity in Section~\ref{sec:fast}, including a
comparison with analytic models. In Section~\ref{sec:twophase}, we
then describe our results for simulations with fast and slow
(`two-phase') winds, followed in Section~\ref{sec:chandra} by an
analysis of fiducial {\em Chandra} observations of our simulated
systems. We give a detailed discussion of our results in
Section~\ref{sec:discussion}, and conclude in
Section~\ref{sec:conclusions}.

\section{The Numerical Method}\label{sec:method}

We use the SPH/$N$-body code {\small GADGET-2} \citep{Springel01,
Springel05} to simulate the dynamics of stars and gas in the
(Newtonian) gravitational field of \sgra. This code solves for the gas
hydrodynamics via the smoothed particle hydrodynamics (SPH) formalism.
The hydrodynamic treatment of the gas includes adiabatic processes,
artificial viscosity to resolve shocks, and optically thin
radiative cooling with the cooling function $\Lambda \approx 6.0
\times 10^{-23} (T/10^7\,$K$)^{-0.7}\,$erg s$^{-1}\,$cm$^{-3}$
\citep{Sutherland93}.

The SMBH is modelled here as a heavy collisionless particle with
gravitational smoothing length of $0.01''$. In addition it acts as a
sink particle: gas passing within the inner radius $R_{\rm in}$
disappears, giving up its mass and momentum to the black hole
\citep{Springel04, DiMatteo05}. The inner radius in our model is a
free parameter, but for most tests we pick it to {  be  $\simlt
0.1''$, much smaller} than the capture, or Bondi radius, $R_{\rm capt} =
G M_{\rm BH}/ (c_{\rm s}^2 + v_{\rm w}^2)$, where $c_{\rm s}$ is the
gas sound speed and $G$ the gravitational constant.  For the case of
\sgra, X-ray observations yield $R_{\rm capt} \sim 1''$
\citep{Baganoff03a}.

At least 99\% of the `wind' (gas) particles escape from the region of
interest, the inner $\sim 10''$, into the Galaxy. Following
these particles in a region which we do not model properly here
becomes prohibitively expensive, and is of limited interest for the
problem at hand. Therefore we eliminate SPH particles that reach an
`outer radius' $R_{\rm out}$. We found that setting $R_{\rm out}$
larger than the distance from the outermost wind source to the SMBH is
appropriate for our purposes.

Stellar winds are modelled via `emission' of new gas (SPH) particles
by the star particles from which these winds emanate. Ejection of
particles is done typically in bursts of about 30 particles per star
and occurs in time intervals of 0.2--1 yrs. For each group of new SPH
particles a random isotropic velocity distribution is generated.  For
tests with moving stars, the full initial particle velocity is the sum
of its random isotropic part (in magnitude equal to the specified wind
velocity $v_{\rm w}$) and the stellar 3-D velocity. The new particles
are given initial temperature of $10^4$ K, which is also the minimum
temperature the gas is allowed to have in the simulations.  As we are
not interested in resolving the wind structure close to stellar
surface, we also give the particles small spatial `kicks' along their
velocities. Baring this, we would have to resolve extremely small
scale structures around stellar surfaces which is not feasible
numerically.  The rest of the SPH properties (pressure, density,
entropy, etc.) are then calculated and followed self-consistently by
the code.

\section{Models with fast winds}\label{sec:fast}

A realistic modelling of \sgra\ wind accretion requires several steps
of varying complexity and importance for final results.  In this
section we shall start by considering the `single-phase' case when the
properties of all the wind sources are identical.  We are interested
in the dependence of the results on the number of sources, their
distribution and their orbital motion, as well as on the choice of our
free parameter $R_{\rm in}$.

\subsection{Fixed stars}\label{sec:fixed}

A very useful test for our numerical methods is the comparison with
analytical results. \cite{Quataert04} presented an analytical 1-D
(spherical) model for the hot gas in the Galactic Centre. He
distributed wind sources in a spherical shell between radii
$2$--$10''$ with stellar number density $n_*(r) \propto r^{-\eta}$,
where $\eta$ is a free parameter varying between 0 and 3. The winds
are further assumed to shock and be thermalized locally.  Essentially,
this model is a spherical wind/Bondi accretion model with distributed
wind sources, and as such it presents a very convenient test bed for
our numerical methods.

In the context of our numerical models, the `on the spot'
thermalization of winds is identical to having an infinite number of
wind sources, since then the mean distance between them is zero and
winds are thermalized immediately.  In practice, we ran simulations
with 200 stars isotropically and randomly distributed in the same
range of radii, with a density profile given by the $\eta=2$
power-law. Our first objective is to test the sensitivity of the
results to different values for the inner radius, $R_{\rm in}=0.1$, $0.4$,
$1''$. The outer boundary is set at $R_{\rm out}=15''$ for these tests. We
ran tests with larger values for $R_{\rm out}$ and found that the results
depend little on this value as long as it includes all the wind
sources.  The total mass loss rate of stars is set to $\dot M_{\rm w} =
10^{-3} \msun\, $yr$^{-1}$, and the wind velocity is $v_{\rm w} =
1000\,$km$\,$s$^{-1}$.

\begin{figure}
\centerline{\psfig{file=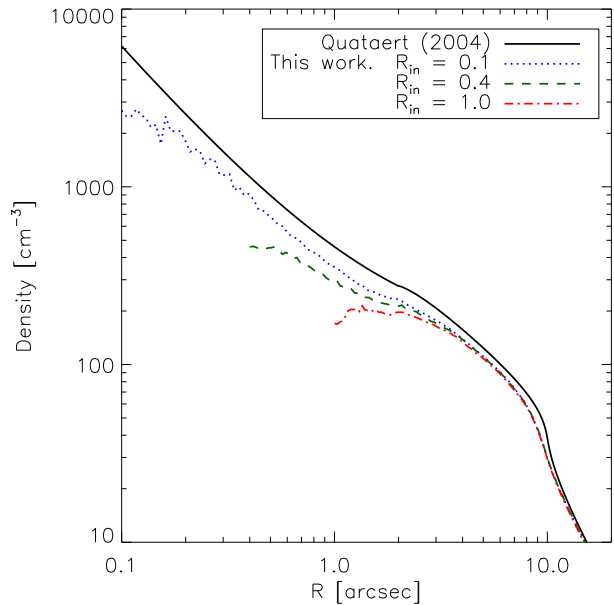,width=.49\textwidth,angle=90}}
\caption{Radial density profiles of the gas in simulations with 200
fixed stars, distributed isotropically. The black solid curve shows
the model of \cite{Quataert04}. Note that our curves (coloured lines)
reproduce his solution rather well down to about twice $R_{\rm
in}$. The values of $R_{\rm in}$ for the different simulations are
given in the inset.}
\label{fig:ndenQ}
\end{figure}

We ran these simulations until time $t \sim 2100\,$yr, at which the
system is in a quasi-steady state. The number of SPH particles in the
steady state is $\simeq 10^6$. To improve statistics, the radial
profiles of quantities of interest are averaged over the last 10
snapshots covering the time interval $t \approx 1800$ --
$2100\,$yr. The resulting density profiles are shown in
Fig. \ref{fig:ndenQ}, together with the result from \cite{Quataert04}
(the model with $\eta=2$)\footnote{We multiplied \cite{Quataert04}'s
density by a factor 2, since his models were actually computed for
$\dot M_{\rm w} = 5 \times 10^{-4} \msun\, $yr$^{-1}$ (Quataert,
priv. comm.).}. Except for the flattening at radii approaching $R_{\rm
in}$, where the SPH particle density is under-estimated due to our
`capture all' boundary condition there, the curves are in good
agreement.

\begin{figure}
\centerline{\psfig{file=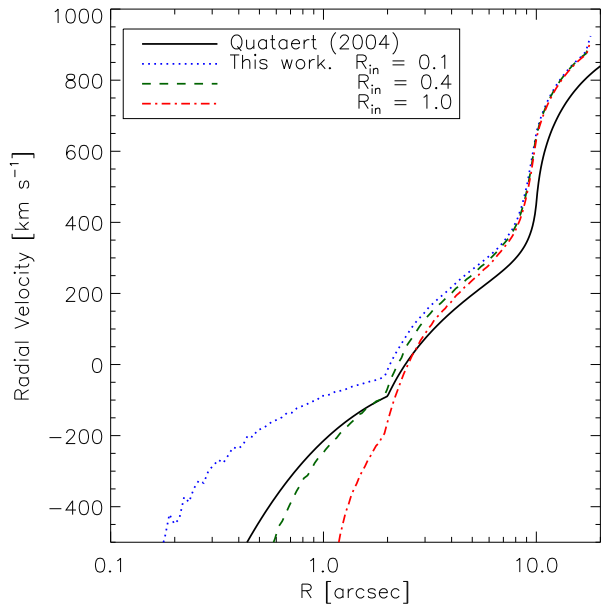,width=.49\textwidth,angle=90}}
\caption{Radial velocity profiles of the gas in the calculations with
200 fixed stars distributed isotropically. Note that for larger
capture radius the gas inflows faster. Because of the finite number of
sources, the outflow velocity of our simulations is slightly larger
than that in \cite{Quataert04}.}
\label{fig:velQ}
\end{figure}

Figure \ref{fig:velQ} shows the average radial velocity as a function
of radius for the same runs. The radial velocity curves differ more
from the \cite{Quataert04} curve (solid line) than the density curves.
Our vacuum inner boundary condition forces the gas next to $R_{\rm
in}$ to inflow with velocity approaching the gas sound speed. Clearly,
when $R_{\rm in}$ is large, this inflow velocity is larger than that
of the \cite{Bondi52} solution at that point. Hence simulations with
large values of $R_{\rm in}$ will over-estimate the accretion rate on
to the SMBH (we discuss this point in more detail below). If instead
$R_{\rm in}$ is a factor of at least several smaller than $R_{\rm
capt}$, we expect to closely match \cite{Quataert04}'s results because
by that point the inflow speed of the \cite{Bondi52} solution is
approaching the local sound speed. This should be achieved in the
simulation with the smallest inner boundary ($R_{\rm in} = 0.1''$).
While we do obtain very similar gas densities (Fig. \ref{fig:ndenQ}),
the inflow velocity is significantly lower in our simulation than in
the analytical solution. We find that, even with 200 sources, the
accretion flow is still strongly anisotropic in the sub-arcsecond
region. Indeed, most of the wind is unbound, with as little as $\sim
1$\% accreting on to \sgra. Thus, out of 200 sources, only the $\sim
10$ innermost stars are important for accretion. This can be seen
through the gas velocity profile: the radial velocity changes its sign
just above the inner radius of the wind-source region.  In addition,
as will become clear later, the incomplete thermalization of the winds
does lead to some excess energy in the wind, and thus {\em not all} of
the gas within the nominally defined accretion radius is actually
bound. Some of the gas at $R < R_{\rm capt}$ may therefore have a
positive radial velocity.

\begin{figure}
\centerline{\psfig{file=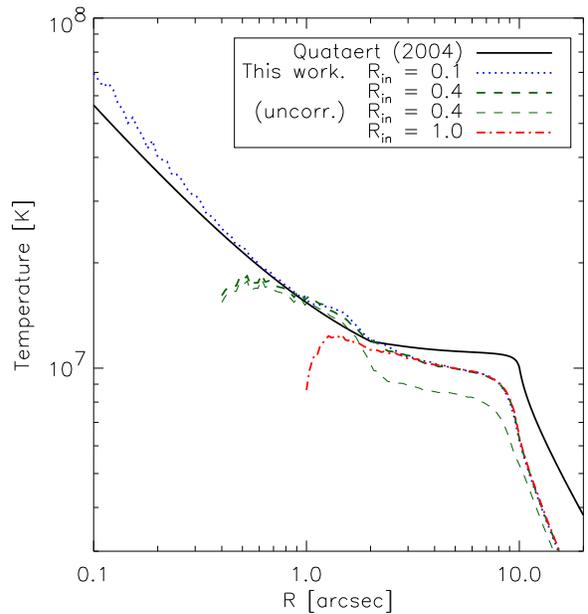,width=.49\textwidth,angle=90}}
\caption{The `1-d' temperature profiles of gas for the simulation with
200 fixed, isotropically distributed stars. The definition of $T_{\rm
1d}$ includes gas bulk motions that would be absent in the case of
exact spherical symmetry; $T_{\rm 1d}$ is to be compared with the
semi-analytical curve of \cite[solid]{Quataert04}.  The `raw' gas
temperature, uncorrected for the bulk gas motions, is given for the
case of $R_{\rm in} = 1''$ (lower red curve).}
\label{fig:tempQ}
\end{figure}

Effects that are due to the finite number of stars should also
manifest themselves in cooler temperature profiles at large radii
compared to the \cite{Quataert04} results. There is a finite distance
the winds will travel before they will experience a shock that will
heat the gas up to the expected temperature. Therefore, for comparison
purposes, we also defined an `effective 1-dimensional' temperature
$T_{\rm 1d} = T + (m_{\rm p} v_{\rm nr}^2 )/(10 k_{\rm B})$, where
$m_{\rm p}$ is the proton mass and $k_{\rm B}$ is the Boltzmann
constant. The quantity $\vec{v}_{\rm nr} = \vec{v} - \langle \vert
\vec{v} \cdot \hat\vec{r} \vert \rangle \hat \vec{r} $ is the local
gas velocity minus the mean radial velocity at the given radius. In a
strictly spherically symmetric model (with an infinite number of
stars) the non-radial velocity $\vec{v}_{\rm nr}$ would of course be
zero, with the corresponding energy converted into thermal energy of
the gas. Therefore $T_{\rm 1d}$ is the quantity to compare with the
temperature derived by \cite{Quataert04}. Figure \ref{fig:tempQ} shows
radially averaged $T_{\rm 1d}$ profiles, and, for the case $R_{\rm in}
= 0.4''$, we also show the actual uncorrected gas temperature $T$
(lower green curve), also averaged in radial shells.  Note that the
difference between $T_{\rm 1d}$ and $T$ is only significant at radii
greater than $1''$ because inside this radius the gas is relatively
well mixed, i.e. shocked.  On the contrary, the average gas
temperature in the outflow region (close to our outer boundary
condition) is significantly smaller than the spherically-symmetric
limit.  The reason for this is that some of the wind from the
outermost stars is actually never shocked as it escapes from the
computational domain.  The same reason is also responsible for the
differences in the radial velocity curves (Fig.  \ref{fig:velQ}): all
of our solutions outflow slightly faster at $R\simlt R_{\rm out}$ than
do the winds of \cite{Quataert04}.

The accretion rate, $\dot M_{\rm BH}$, in our simulations is defined
as the total mass entering the sphere of radius $R_{\rm in}$ per unit
time.  We find that the accretion rate increases initially while winds
fill in the available space, but after less than 1,000 years $\dot
M_{\rm BH}$ reaches a steady state value.  The final value will of
course depend on the choice of $R_{\rm in}$. For the 
simulations described above, we get accretion rates $\dot M_{\rm BH}
\approx 1.5$, $3.5$ and $7 \times 10^{-5} \msun\,$ yr$^{-1}$, for
$R_{\rm in} = 0.1, 0.4,\,\, {\rm and}\; 1''$ respectively. The
increase of $\dot M_{\rm BH}$ with $R_{\rm in}$ implies that a
significant fraction of the gas that is `accreted' in the simulations
with large $R_{\rm in}$ would not have done so had we been able to
resolve the smaller scale flow.  As far as the exact accretion rate
values are concerned, smaller $R_{\rm in}$ is better, but, since a
smaller $R_{\rm in}$ requires shorter integration steps, one has to
make a pragmatic compromise and choose a value of $R_{\rm in}$ that
allows simulations to be run in a reasonable time.

The accretion rates that we obtain are comparable with the
\cite{Quataert04} result for $\eta=2$, $4.5\times 10^{-5} \msun\,$
yr$^{-1}$. However, the most reliable of the tests, with $R_{\rm
in}=0.1''$, shows an accretion rate a factor of 3 lower than the
semi-analytical result. We interpret this as another manifestation of
the incomplete spherical symmetry due to the finite number of stars in
our simulations. We found that even at radii as small as $0.3''$ there
are large deviations from the mean in the gas velocity in the same
radial shell. In addition, a fraction of the gas has significant
specific angular momentum. To test these points further, we ran an
additional simulation with $R_{\rm in} = 0.1''$, but with 40 wind
sources, which is more realistic as far as \sgra\ wind accretion is
concerned but should enhance the discreteness effects when compared
with \citet{Quataert04}. The results we obtain are in general similar,
but the effects produced by the finite number of sources are indeed
enhanced.  For example, the density in the inner region is $\sim 50\%$
lower than in the simulation with 200 stars, producing a corresponding
lower accretion rate (see green curves in Figs. \ref{fig:ndenrot} and
\ref{fig:accraterot}).

We find that our derived accretion rate values are about one order of
magnitude lower than those of \cite{Coker97}. The differences are
however due to a different physical setup rather than
numerics. \cite{Coker97} used $v_{\rm w} = 700$ km s$^{-1}$, whereas
we used $v_{\rm w} = 1000$ km s$^{-1}$ here. This difference in wind
velocity alone accounts for a factor of $\approx 3$ difference in
$\dot M_{\rm BH}$. Furthermore, their total mass loss rate from the
young stars is higher than ours by a factor of 3. If we take into
account these differences, our simulation with the smallest capture
radius and 40 fixed stars appears to be consistent with the results of
\cite{Coker97}. Finally, the work of \cite{Rockefeller04} did not
directly focus on the accretion on to \sgra. They were mainly
interested in the dynamics of gas on parsec-scales. The accretion rate
they quote is likely to be over-estimated because of the large value
of $R_{\rm in}\approx 2''$ they use.

\subsection{Stars in orbits around \sgra}\label{sec:moving}

The next step towards a more realistic model is to allow the stars to
follow orbits in \sgra's gravitational potential. We do this in two
idealised configurations: a spherically symmetric and a disc-like
system (for the latter, the stellar disc is somewhat thick, with
$H/R\sim 0.2$, as seems to be the case observationally). In both
cases, we put 40 stars in circular orbits around the SMBH. We use the
same stellar density profile as in Section~\ref{sec:fixed}, $n_*(r)
\propto r^{-2}$, and boundary conditions $R_{\rm in} = 0.1''$, $R_{\rm
out}=18''$.

\begin{figure}
\centerline{\psfig{file=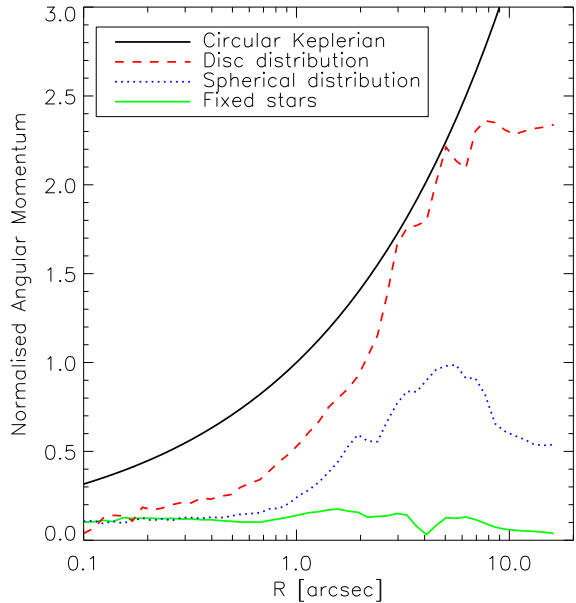,width=.49\textwidth,angle=90}}
\caption{Angular momentum averaged on radial shells, $l(r)$, in the
simulations with 40 stars in circular orbits (red and blue lines) and
in fixed positions (green). For comparison we also show the value of
the angular momentum of a circular orbit (black).}
\label{fig:angmom}
\end{figure}

The significant difference with respect to the simulations discussed
in Section~\ref{sec:fixed} is that now the gas particles have a
significant net angular momentum. To quantify this we create radial
profiles of the average angular momentum, defined as the modulus of
the average angular momentum vector in a shell, $l(r) = \vert \langle
\vec{r} \times \vec{v} \rangle \vert$. Since we measure distances in
arc-seconds and velocities in units of circular Keplerian velocity at that
distance, a circular orbit at $r=1''$ has $l(1'') = 1$ in these
units. Note that $l(r)$ should vanish for an isotropic orbital
distribution of wind sources, even if each individual gas particle has
a high angular momentum. Figure \ref{fig:angmom} shows the $l(r)$
profiles for the simulations described in this section, and for the
one with 40 fixed stars from Section~\ref{sec:fixed}.

When the stars are confined to a disc (red curve), the gas has on
average roughly the local Keplerian angular momentum at the `wind
source' region, simply because the stars are on Keplerian circular
orbits with the same angular momenta direction. If the orbits are
randomly oriented (blue), the angular momenta of the gas should cancel
out owing to the symmetry.  However, the cancellation is incomplete
due to the finite number of sources. Finally, in the case where the
stars are fixed (green), the angular momentum is negligible, as it
should.

Comparing the angular momentum in the three simulations in the
sub-arcsecond region, we see that all simulations yield similar
results. In all the cases the gas in this region is rotating
significantly slower than the local Keplerian rotation, indicating
that centrifugal support is not important for the gas. This is
somewhat surprising given the vastly different angular momentum curves
at larger radii and is not a result of viscous transport processes,
because a physical viscosity was not included in the
simulations\footnote{Note that the fact that the gas is significantly
sub-Keplerian in the region justifies our neglect of angular momentum
transport by viscosity.}.  This result rather seems to indicate that
only the gas with originally low enough angular momentum makes it into
the innermost region and is subsequently accreted, {  as already
discussed by \cite{Coker97}}. The fraction of the stellar wind with
low angular momentum however varies greatly between the simulations,
which in turn explains the different accretion rates.

\begin{figure}
\centerline{\psfig{file=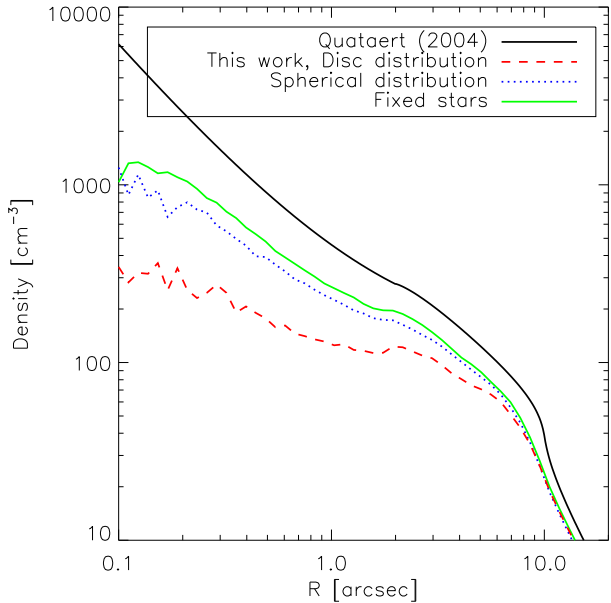,width=.49\textwidth,angle=90}}
\caption{Density profiles of the gas in the calculations with 40 stars
for different orbital configurations. The presence of significant
angular momentum prevents the gas from inflowing, resulting in a lower
density in the inner part.}
\label{fig:ndenrot}
\end{figure}

Note that, once again due to the finite number of stellar sources,
there is a distribution in the values of the gas angular momentum for
any given radius. Some of the gas at $R\sim 0.5''$, for example, has a
roughly Keplerian angular momentum which prevents it from
accreting. This gives rise to shallower density profiles, as seen in
Fig.~\ref{fig:ndenrot}, and is particularly important in the disc
simulation (red curve), where the density is 2--3 times lower at $1''$
than in the case with fixed stars. Note that in this case, the density
value at $1''$ is in better agreement with densities implied by
{\em Chandra} observations \citep{Baganoff03a}. Correspondingly, the
mass supply to the central black hole also decreases: we find that in
the simulation with stars located in a disc, the average accretion
rate is only $\sim 2 \times 10^{-6} \msun\,$ yr$^{-1}$, about 5 times
lower than the value found with fixed stars.

\begin{figure}
\centerline{\psfig{file=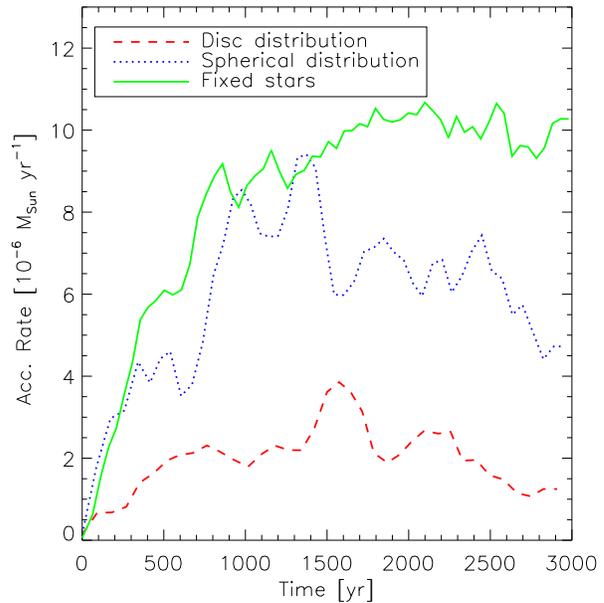,width=.49\textwidth,angle=0}}
\caption{Accretion rate as a function of time in the simulations with
40 stars for different orbital configurations. The rotation of the
stars produces a lower but variable accretion rate.}
\label{fig:accraterot}
\end{figure}

In addition to these effects, stellar motions yield a significant
variability in the accretion rate. Figure \ref{fig:accraterot} shows
the accretion rate as a function of time for the same three
simulations. In all the simulations, the accretion rate increases
initially, as the stellar winds fill up the space. In the run with
fixed stars, the accretion rate is practically constant after this
initial increase. In contrast, when stars are allowed to follow
orbits, the geometry of the stellar system changes with time (both in
the isotropic and disk-like configurations), and so does the fraction
of gas that can flow to the inner region. We should emphasise that
this factor of $\sim 30-70$\% variability occurs in the two rather
simple and still idealised stellar wind systems. In the more realistic
situation, the variability would be enhanced further because we expect
(i) yet smaller number of wind sources, (ii) a more diverse set of
stars with different mass loss rates, (iii) non-circular stellar
orbits, (iv) intrinsic stellar wind variability for luminous blue
variable stars, etc.

\section{Fast and slow (two-phase) winds  } \label{sec:twophase}

The next ingredient we add to our treatment of accretion of stellar
winds on to \sgra\ is the presence of slow winds. Previous models have
considered only wind velocities $\simgt 600$ km s$^{-1}$, as measured
by \cite{Najarro97}. However, \cite{Paumard01} showed that several of
the inner stars with important mass-loss rates emit winds with
velocities as low as $\sim 200$ km s$^{-1}$. This has important
consequences, because the slow winds are expected to cool and form
clumps \citep{C05}.

We use 20 wind sources in total, with each star loosing mass at the
rate of $\dot M_* = 4 \times 10^{-5} \msun$ year$^{-1}$.  To resemble
the two observed populations of luminous blue variable candidates,
`LBVs', and Wolf--Rayet stars, `WRs', we assume that stars in our
simulations can have either $v_{\rm w} = 300\,{\rm km\, s}^{-1}$ or
$v_{\rm w} = 1000$ km s$^{-1}$, characteristic values for the two
populations respectively. The values for the inner and outer boundary
conditions are $R_{\rm in} = 0.07''$ and $R_{\rm out} = 9''$.  We run
the simulations for $\approx 4,000\,$yr, then the number of particles
reaches $N_{\rm SPH} \approx 1.5 \times 10^6$.

For the distribution of the stars we adopt two different
configurations. First, we place the stars into two discs in order to
reproduce the distribution reported by \cite{Genzel03a}. We also place
most of the LBVs in the inner disc, following
\cite{Paumard01}. However, the stellar orbits are still only roughly
known and the latest observations suggest that the LBVs may actually
be distributed more evenly among both stellar systems (Genzel,
priv. comm.). For these reasons, we consider a second simulation where
the stars are distributed isotropically. The real distribution of the
mass-loosing stars should be somewhere in between these two extreme
cases.

The simulations presented here differ from the work of \cite{C05} in
that we do not split particles that get relatively close to the
SMBH. The splitting was used to increase the resolution in the inner
part. However, further tests showed that some loss of accuracy can
occur in our present implementation of this approach, so that it
failed to really show a convincing practical advantage.

\subsection{Stellar wind sources placed in two discs}\label{sec:2discs}

Here we attempt to model the \sgra--stellar wind system by setting up
the stellar source distribution in a way that resembles the
observations of the inner parsec in the GC \citep{Paumard01,
Genzel03a}. Our approach is the same as the one described by
\cite{C05}, therefore we describe it only briefly here.

The stars are distributed uniformly in radius and in two perpendicular
rings. All stars follow circular Keplerian orbits. The radial extent
of the rings is from $2''$ to $5''$ and $4''$--$8''$ for the inner and
outer one respectively.  Out of the 20 wind sources, we assume that 6
LBV-type and 2 WR-type stars are in the inner ring, and the outer ring
is populated by 3 LBVs and 9 WRs.

\subsubsection{Large scale structure of the gas flow}\label{sec:large}

\begin{figure*}
\begin{minipage}[b]{.49\textwidth}
\centerline{\psfig{file=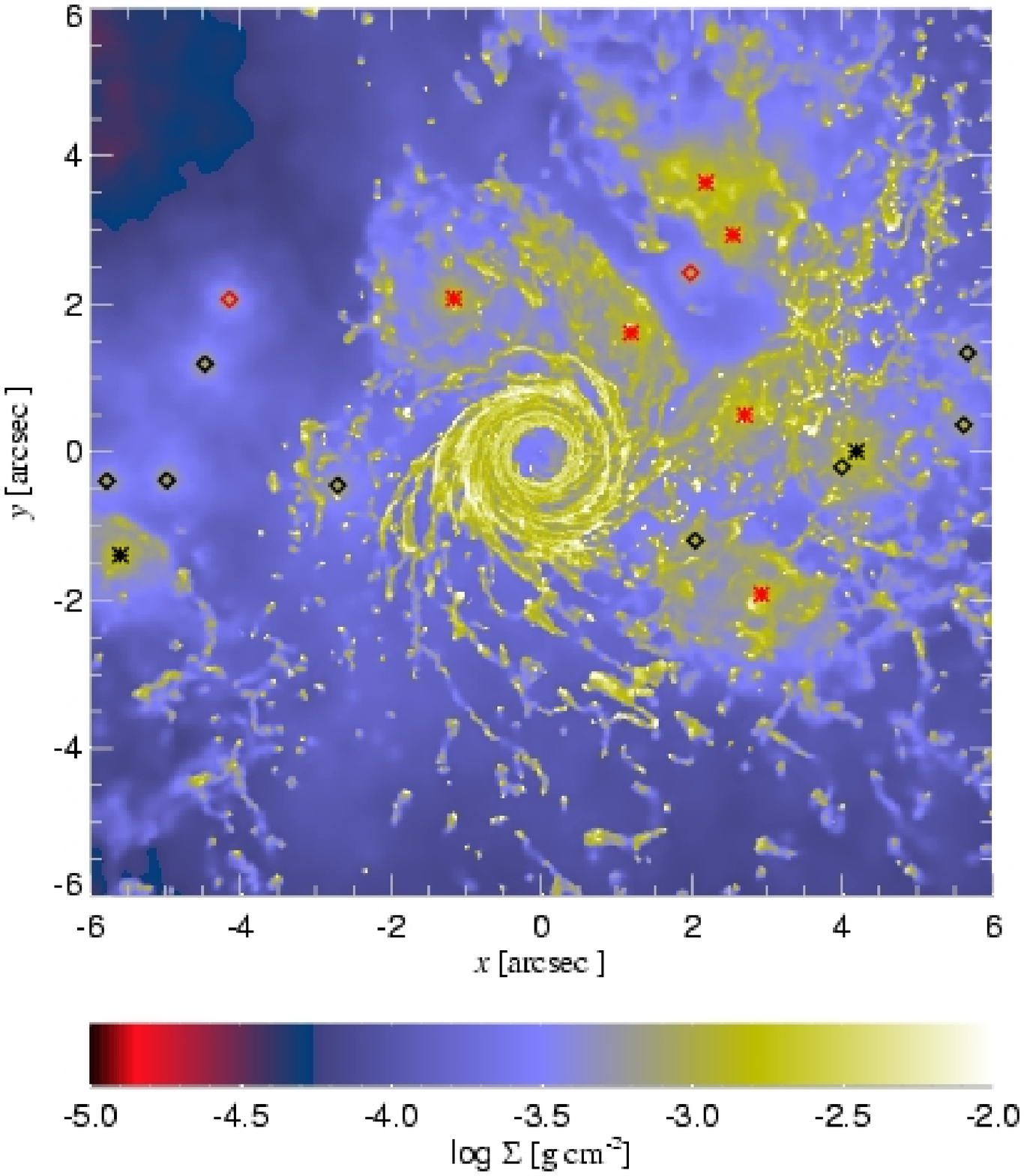,width=.99\textwidth,angle=0}}
\end{minipage}
\begin{minipage}[b]{.49\textwidth}
\centerline{\psfig{file=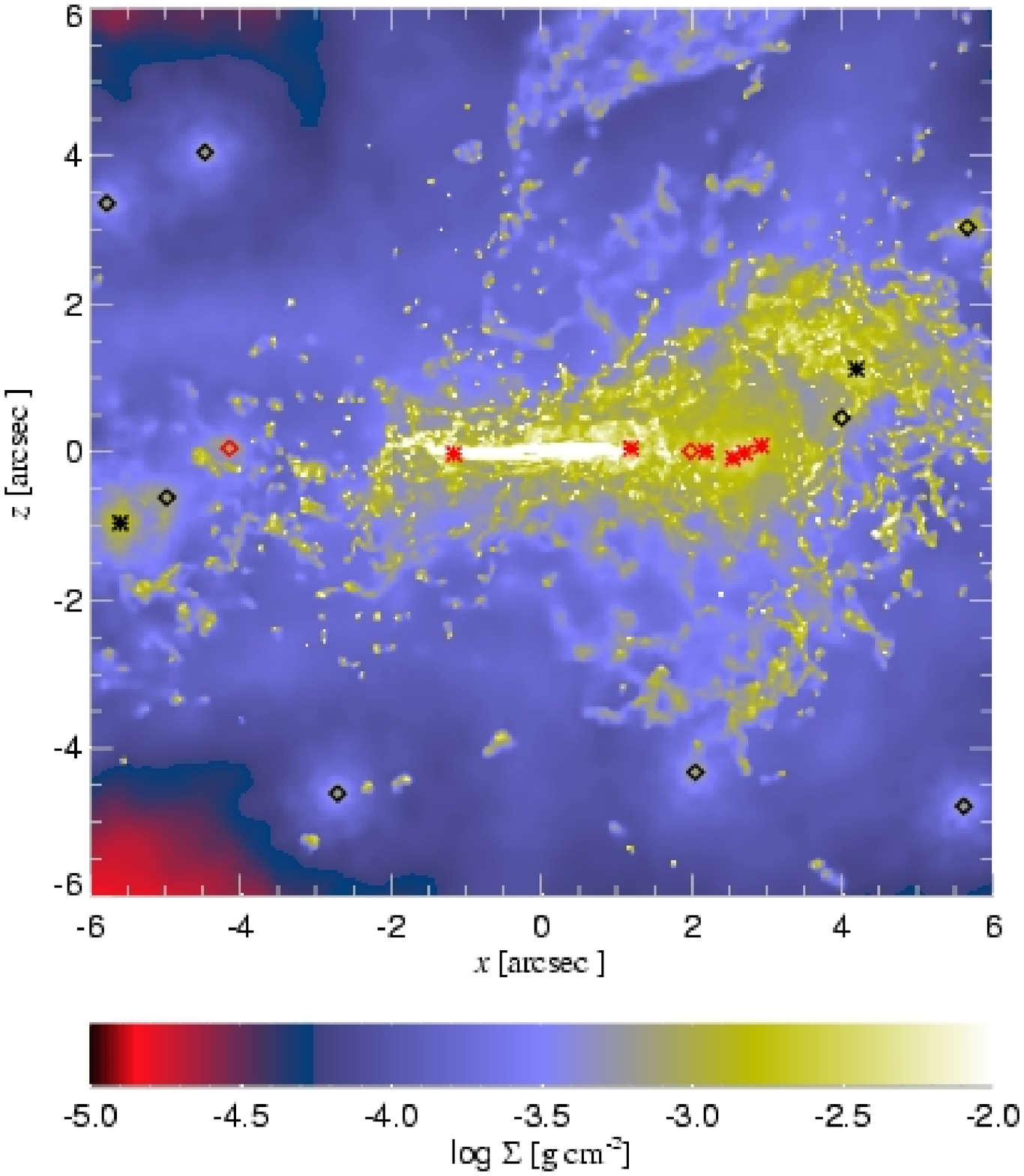,width=.99\textwidth,angle=0}}
\end{minipage}
\caption{Left panel: Column density of gas in the inner $6''$ of the
computational domain 2,450 years after the beginning of the
simulation. Stars of the inner disc are shown as red symbols, while
the ones on the outer disc appear in black. Asterisks mark slow wind
stars (LBVs), whereas diamonds mark stars producing fast winds
(WRs). The inner stellar ring is face-on while the outer one is seen
edge-on (plane $y=0$). Right panel: same as on the left, but the outer
stellar ring is now seen face-on while the inner one appears edge-on
($z=0$).}
\label{fig:largeview}
\end{figure*}

Figure \ref{fig:largeview} shows the resulting morphology of the gas
2.5 thousand years into the simulation. The inner ring, mainly
composed of LBVs that are shown by asterisks, is viewed face-on in the
left panel, and edge-on in the right one. The inner ring stars are
shown with the red coloured symbols whereas the outer disc stars are
painted in black. The inner stars are rotating clock-wise in the left
panel of the figure, and the outer ones rotate counter clock-wise in
the right panel.

Cool dense regions in the gas distribution (shown as bright yellow
regions in Fig. \ref{fig:largeview}), are mainly produced by winds
from LBVs.  When shocked, these winds attain a temperature of only
around $10^6$ K, and, given the high pressure environment of the inner
parsec of the GC, quickly cool radiatively \citep{C05}. LBV winds form
bound clouds of gas, often flattened into filaments due to the SMBH
potential and the symmetry of the problem. As more filaments are
formed in the inner region, they start overlapping and eventually form
a disc that lies almost at the plane $z=0$.  The orientation of the
disc plane is very close to that of the inner stellar ring, which at
first may appear surprising given that there are 3 other LBV stars at
the larger stellar disc. However, the escape velocity from the outer
ring is of the order of the stellar winds velocity for the LBVs, and thus a
large fraction of the slow wind from the outer ring escapes the
computational domain and never influences the inner cold disc
orientation.

The fast winds contribute to the inhomogeneity of the cold gas. This
is well illustrated by one of the two WR-stars placed in the inner
ring at $(x,y) \approx (2, 2.5)$ in the right panel of
Fig.~\ref{fig:largeview}. The wind from this star has more mechanical
power than all of the winds from the other LBVs in the simulation
($L_{\rm mech} \propto \dot M_* v_{\rm w}^2$). It cuts the combined
wind of four neighbouring LBVs into two bands of gas. These two
streams of gas are further stretched out by tidal shear from the SMBH
and then chopped into smaller cool blobs by the interaction with the
winds of the outer ring.

The three LBVs placed in the outer ring add complexity to the
morphology of the cool gas as they produce cool gas streamers moving
in a plane different from that of the inner one. One of these stars is
at $(x,z)\approx (-5.5, -1)$. It produces most of the cool blobs of
gas seen at the left of the right panel. Also note that the LBV at
$(x,z)\approx (4, 1)$ is uplifting  a significant amount of the cooled
wind from the inner LBVs. On the other hand, the WRs by themselves do
not produce much structure, as seen for instance from the two stars
located at the upper left corner on the right panel. The fast winds
they produce have temperatures $\simgt 10^7\,$K after shocking, and do
not cool fast enough to form filaments.

\begin{figure*}
\begin{minipage}[b]{.49\textwidth}
\centerline{\psfig{file=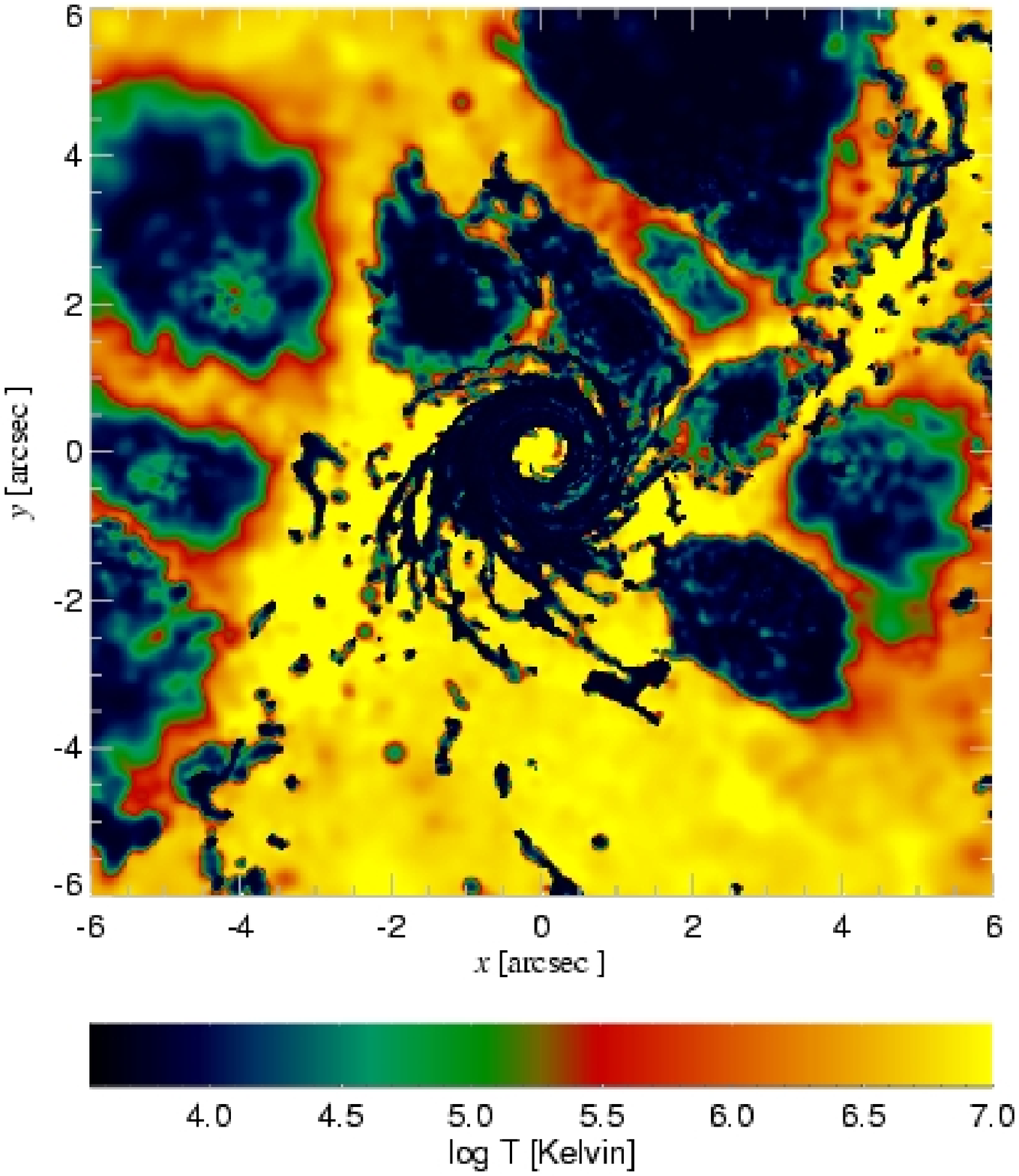,width=.99\textwidth,angle=0}}
\end{minipage}
\begin{minipage}[b]{.49\textwidth}
\centerline{\psfig{file=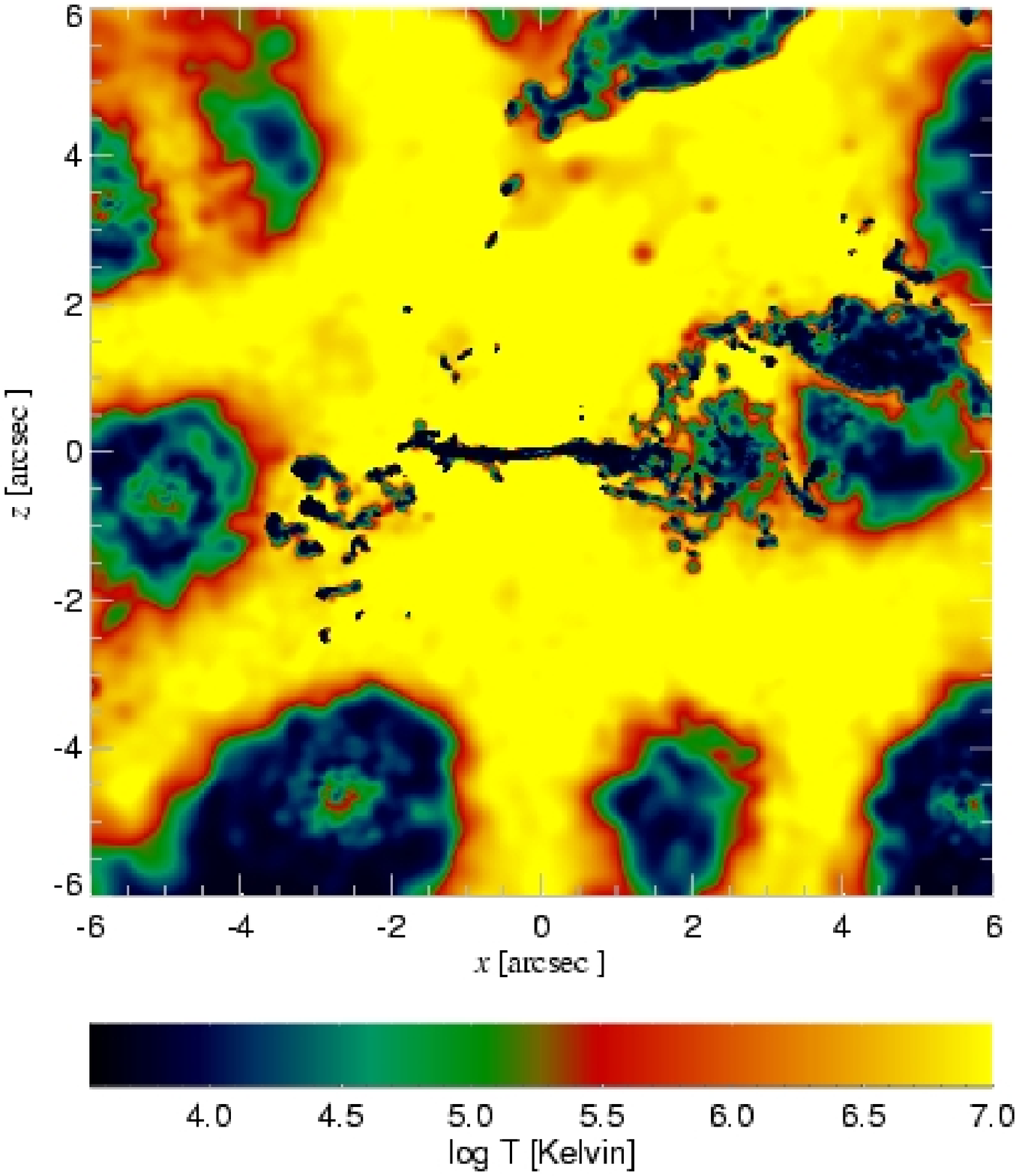,width=.99\textwidth,angle=0}}
\end{minipage}
\caption{Similar to Fig. \ref{fig:largeview}, but showing $1''$-thick
cuts of the gas temperature. Stars are not shown for clarity.  The
left panel shows a slice between $z = -0.5''$ and $+0.5''$, whereas
the right one shows the gas temperature in the range $-0.5'' < y <
+0.5''$. The minimum temperature in the simulation is set to $10^4$
K. In reality gas would cool even further, likely to a few hundred K.}
\label{fig:largetemp}
\end{figure*}

Figure \ref{fig:largetemp} shows the average gas temperature in a
slice $z=\pm 0.5''$ (left panel) and in a slice $y=\pm0.5''$ (right
panel) of the inner $6''$ region\footnote{To avoid low-density hot
regions from dominating the map, we column-averaged $\log T$ rather
than simply $T$.}.  The clumps and the disc are cold as a result of
radiative cooling, and before they collide the stellar winds have a
low temperature as well. However, close to the stars the temperature
appears higher than it would have been in reality. This is an artifact
of the injection of a finite number of SPH particles per time-step
into the wind, which creates some small-scale structure in the density
and pressure profiles, leading to energy dissipation.  This effect is
however of minor importance since the temperature in these regions is
still relatively low compared to the maximum temperatures that the gas
can attain in a shock. In other words, the energy of the winds is
still strongly dominated by the bulk motion of the gas rather than
thermal energy.

The diffuse gas filling up the rest of the space is hot, with
temperature $\sim 10^7$ K, comparable to that producing X-ray emission
detected by {\em Chandra}.  Gas cooler than that would be invisible in
X-rays due to the finite energy window of {\em Chandra} and the huge
obscuration in the Galactic plane.

\subsubsection{Structure of the inner flow}\label{sec:inner}

\begin{figure}
\centerline{\psfig{file=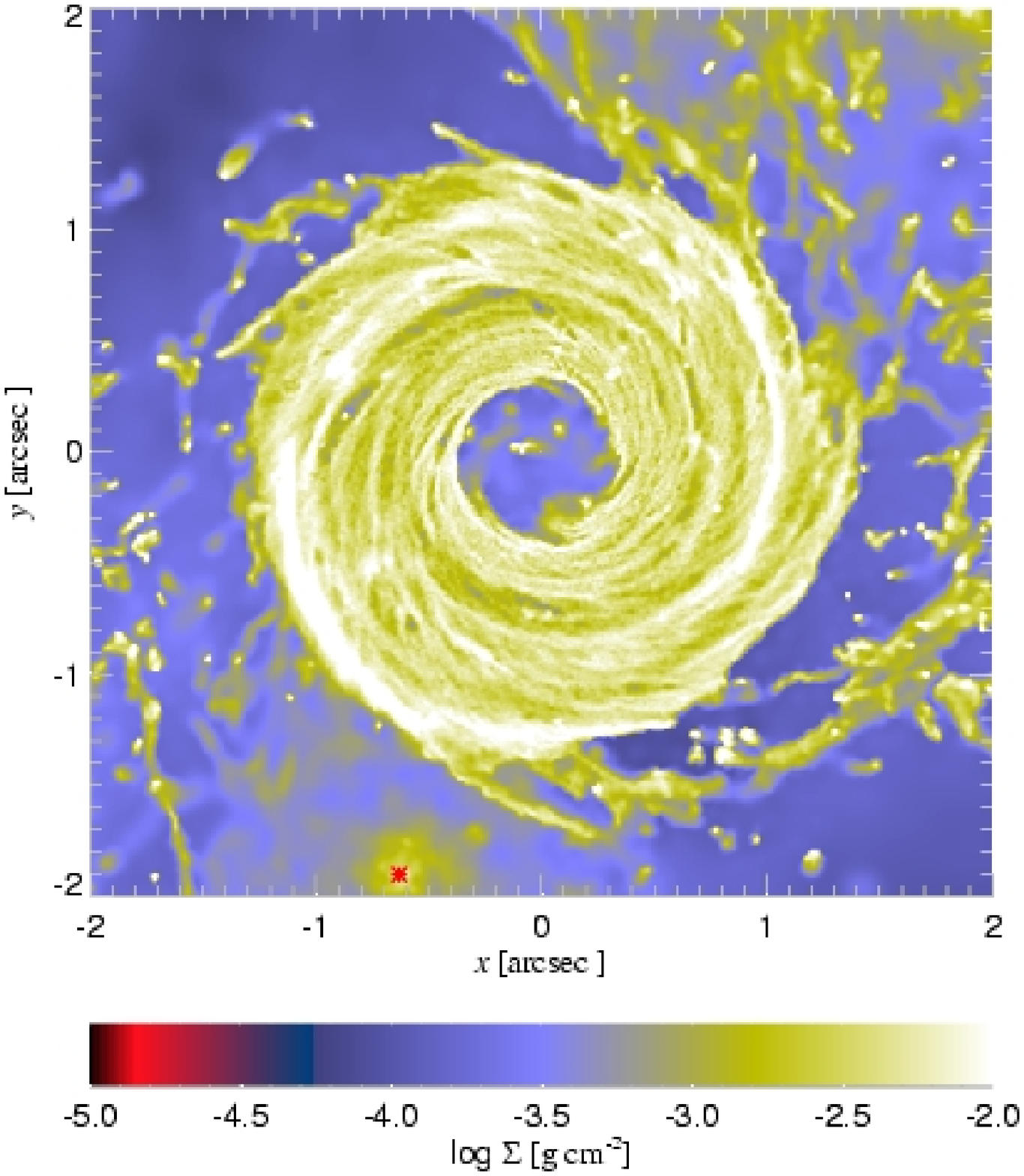,width=.49\textwidth,angle=0}}
\caption{Top view of the inner $2''$ cube at time $t=4.0 \times 10^3$
years. Note that the disc grew larger in radial extent but the inner
region is still devoid of cool gas except for a few clumps.}
\label{fig:dens2}
\end{figure}

Figure \ref{fig:dens2} depicts the structure of the inner flow later
in the simulation, at $t=4.0\times 10^3$ years. At this late time, the
cool disc becomes heavier and larger in the radial direction than seen
in Fig. \ref{fig:largeview}. The inner $\sim 0.3''$ region is still
devoid of cool gas except for a few filaments. This is due to two
reasons. First, the angular momentum of the slow winds from the
innermost stars is not zero, even for the wind directed in the
opposite direction of the stellar motion. The Keplerian circular
velocity at $2''$ is $v_{\rm K} \approx 440$ km/sec, whereas the wind
velocity is $v_{\rm w} = 300$ km/sec. Thus wind with even the minimum
angular momentum would circularise at $r \sim 2'' \times
(v_{\rm K}-v_{\rm w})^2/v_{\rm K}^2 \sim 0.2''$. However, due to interactions of this
gas with the gas with higher angular momentum, the minimum disc radius
is actually a factor of $\sim 2$ higher. Second, the inner empty
region of the disk is not filled in by the radial flow of cold gas
through the disc because the viscous time scale of the cold disc is
enormously long compared with the duration of the simulation
\citep[e.g.][]{NC05}. Gas clumps having low angular momentum infall
into the inner arcsecond, but this does not occur frequently enough to
fill in that region, and most of these clumps appear to be disrupted
and heated in collisions.  The long viscous time-scale is also the
reason why the filaments do not merge forming a smooth disc at larger
radii. The individual filaments that give shape to the disc are still
distinguishable in the figure.

The mass of the cold disc is actually quite small, i.e. only $\approx
0.2 \msun$. It is instructive to compare this number with the mass of
the wind produced by the `LBV' stars in $4\times 10^3$ years, which
turns out to be $1.4 \msun$ for this simulation. Obviously most of the
cold gas escapes from the simulation region. However, some of this gas
does not have a true escape velocity. Had we simulated a larger
region, a fraction of this cold gas could return to the inner region
on highly eccentric orbits, producing further variability and
complexity in the morphology of the cold gas and accretion history of
\sgra.

\subsubsection{Accretion on to \sgra}

\begin{figure}
\centerline{\psfig{file=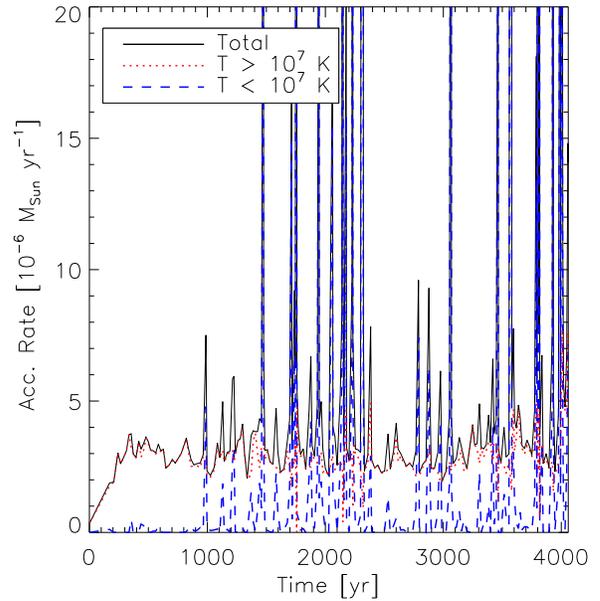,width=.49\textwidth,angle=0}}
\caption{Accretion rate on to the SMBH as a function of time (black
line). This rate is then divided into that of hot gas (red) and that
of low temperature gas (blue). Note that the time-averaged accretion
is dominated by the hot component that is quasi-constant; the
accretion of cold gas is episodic and highly variable, but smaller
overall.}
\label{fig:accratelbvwr}
\end{figure}

As described in Section~\ref{sec:method}, all gas particles entering
the inner boundary of the computational domain, in this case $R_{\rm
in} = 0.07''$, are presumed to be instantaneously accreted by
\sgra. In practice, such accretion would happen on the viscous time
scale of the flow, which depends on the viscosity $\alpha$-parameter \citep{Shakura73},
the gas temperature and the circularisation radius. When taking these
factors into account, we would expect a time-scale of at least 10 yr
for the accretion of hot gas. For this reason, in
Fig. \ref{fig:accratelbvwr}, we plot the accretion rate smoothed over
a time-window of duration $\Delta t \approx 10$ years.

Figure \ref{fig:accratelbvwr} also shows the accretion rate of hot ($T
> 10^7\,$K, red line) and cold ($T < 10^7\,$K, blue line) gas
separately. The accretion rate of hot gas is fairly constant, and has
a value $\dot M_{\rm BH} \approx 3 \times 10^{-6} \msun\, $yr$^{-1}$,
consistent with the {\em Chandra} estimates \citep{Baganoff03a}. In
addition to this component, the intermittent infall of cold clumps
produces most of the variability in the accretion rate. The cold gas
falls into the innermost regions due to two factors:  (i) it is
initially created on low angular momentum trajectories, and/or (ii)
it acquires such orbits through collisions with other cold gas clumps.

\begin{figure}
\centerline{\psfig{file=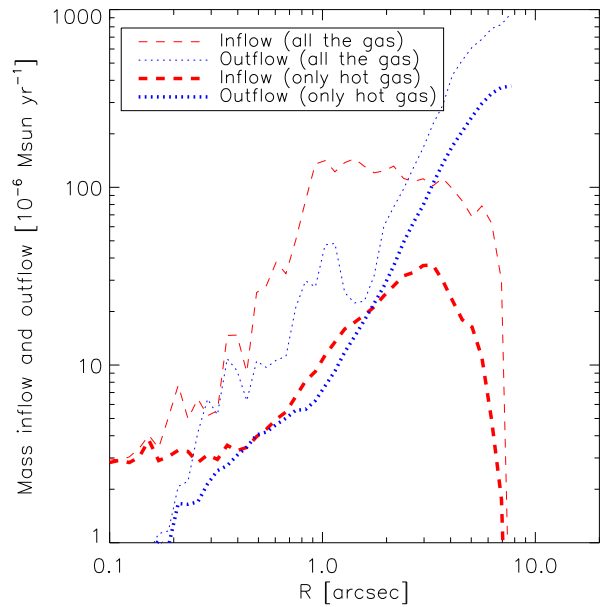,width=.49\textwidth,angle=90}}
\caption{Mass inflow (red curves) and outflow (blue) as a function of
radius. The thick lines show the profiles for the hot gas only, while
the thin lines include all the gas.}
\label{fig:massflow}
\end{figure}

For further analysis we define the local mass inflow and outflow rates
integrated on shells at a given radius. Figure \ref{fig:massflow}
shows both of these quantities, for all the gas (thin lines), and for
the hot gas only ($T > 10^7\,$K, thick lines). The profiles were
measured in a similar fashion as those in Section~\ref{sec:fast} by
stacking 10 snapshots at $t \approx 2.5 \times 10^3\,$yr. For the hot
gas, the outflow practically cancels the inflow between $0.4''$ and
$2''$.  Only the gas with low enough angular momentum and thermal
energy accretes in this region, the rest escapes into the
outflow. This effect is completely absent in a spherically symmetric
flow such as the \cite{Bondi52} solution.

The true accretion flow in our simulation starts at $R\approx 0.4''$,
where the inflow rate becomes approximately constant and the outflow
is negligible. This radius would be a better definition of the
effective capture radius.

The profiles that include all the gas (thin lines) are dominated by
cold gas in the disc at $R \approx 0.3$--$1.3''$. We find that the
mean eccentricity for this gas is $e \approx 0.1$. This deviation from
circular velocity, together with the relatively high mass of the disc
explain the `noise' in the inflow/outflow profiles in this region.
However, the inflow dominates, and its net rate is much larger than
the accretion rate at $R_{\rm in}$, so an important fraction of the gas
actually stays in the disc, making it grow in mass. For larger radii
the inflow becomes negligible and the outflow rate reaches
approximately the total mass loss rate of the wind sources, $\dot M_{\rm w}
= 10^{-3} \msun\, $yr$^{-1}$.

\begin{figure}
\centerline{\psfig{file=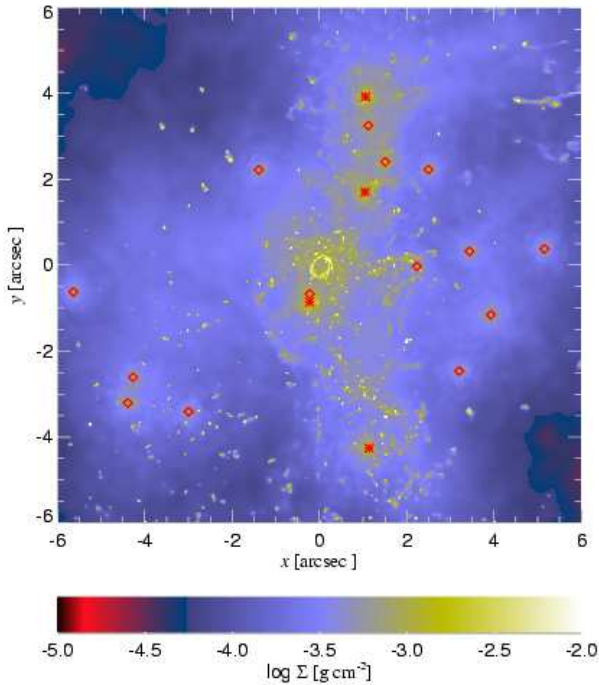,width=.49\textwidth,angle=0}}
\caption{Gas column density map for a simulation with 20 stars, chosen
as in Section~\ref{sec:2discs} but with orbits that have a random
angular momentum vector.}
\label{fig:iso}
\end{figure}

\subsection{Simulation with isotropic orbits}\label{sec:isotropic}

With the aim to test how the results depend on the assumed stellar
orbits we run a simulation with exactly the same set-up as in the
previous subsection, but instead of placing the stars into the two
rings, we oriented the stars' circular orbits randomly. The
distribution of angular momentum vectors is obviously isotropic for
this case.

Figure~\ref{fig:iso} shows the column density of the gas at $t =2.4
\times 10^3\,$ years after the beginning of the simulation. For this
simulation, no conspicuous disc is formed. Instead, we find that only
a small ring on scales of $\sim 0.5''$ is formed. The ring has the
same angular momentum direction than the innermost star (an LBV). Its
radial scale is smaller than the inner boundary of the disc formed in
Section~\ref{sec:inner}, because in this simulation the gas that
formed the ring interacts with a larger quantity of gas with different
angular momenta, loosing its own.

\subsection{Comparison of both simulations}\label{sec:both}

The prominent disc-like structure that we find in the simulation with
two stellar discs appears to be inconsistent with some of the current
observations. In infrared images \citep[e.g.,][] {Scoville03, Paumard04}
complex structures of cold gas are indeed seen, but they are not in a
disc configuration.  In addition, the cold gas structures are
typically observed on larger scales than what we find in our
simulation. The most likely reason for the discrepancies between these
recent observations and our model may be ascribed to the specific
initial conditions we used. In particular, placing many LBVs in the
same plane at a short distance from the SMBH may be particularly
favourable for the development of a disc. Completely different is the
case in which the stellar orbits are oriented randomly (Section~\ref{sec:isotropic}). The distribution of angular momenta vectors is
obviously isotropic for this case, so there is no obvious preference
for a particular plane where a disc would form. The tiny ring formed
there perhaps could be missed observationally or not formed at all if
closest slow wind stars are in reality farther away than we assumed
for the simulation.

Nevertheless, in both simulations we get a two-phase medium in the
inner region, and a comparable accretion rate. These are robust
results, dependent only on the velocities we have chosen. The specific
gas morphology, instead, depends strongly on the orbital geometry of
the initial source distribution.

\begin{figure}
\centerline{\psfig{file=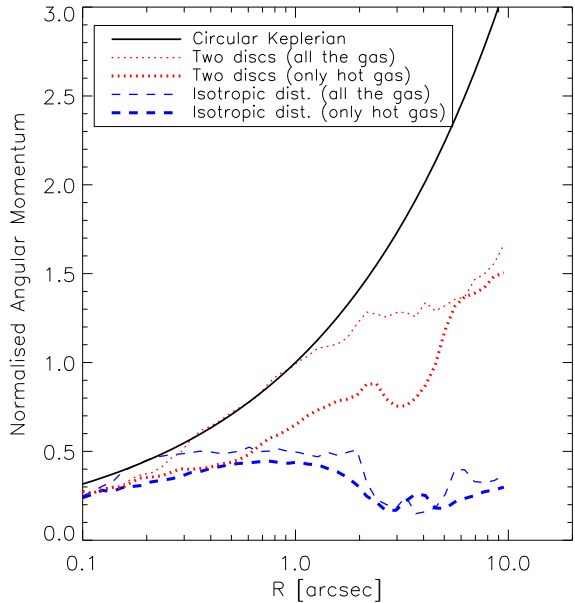,width=.49\textwidth,angle=90}}
\caption{Angular momentum profiles of the simulations with isotropic
orbits (blue, Section~\ref{sec:isotropic}) and with two discs (red,
Section~\ref{sec:2discs}). For comparison, the value of a Keplerian
circular orbit is shown as well (black line).}
\label{fig:angmom_lbv}
\end{figure}

The angular momentum profiles (defined as in Section~\ref{sec:moving})
of gas in the simulations with isotropic orbits (blue) and with two
discs (red) are shown in Fig.~\ref{fig:angmom_lbv}. In the isotropic
case the angular momentum is low at the wind source region. In
contrast, the gas angular momentum is much higher in the two stellar
disc simulations at the same range of radii. Somewhat surprisingly,
the angular momentum content of hot gas in the inner arcsecond is
actually quite similar for the two simulations. This can be traced to
the hydrodynamical interaction of the hot gas with the cold gas, whose
net angular momentum is much higher. Indeed, the angular momentum
profiles of all the gas (cold and hot), shown with thin curves in
Fig.~\ref{fig:angmom_lbv}, are nearly Keplerian in the inner
arcsecond region.

\subsection{Comparison of the two-phase and one phase
simulations}\label{sec:1vs2} 

At this point it is also instructive to highlight some of the
systematic differences between the `one-phase' (fast winds only) and
the `two-phase' simulation results. The angular momentum profiles of
the one-phase and two-phase simulations are shown in
Figs.~\ref{fig:angmom} and \ref{fig:angmom_lbv}, respectively. While
in the one-phase simulations the angular momentum of gas near the
capture radius was always much lower than the local Keplerian value,
the two-phase simulations show almost Keplerian rotation near the
inner boundary. As explained in Section~\ref{sec:both}, this is caused
by the interaction between the hot and the cold gas phases. Only the
part of the fast wind that has a low angular momentum accretes from
the hot phase, because its pressure support against SMBH gravity is
already high. In contrast, cold gas has little pressure support and
hence even gas with a rather high angular momentum can be captured
into the central arcsecond by \sgra.  This also implies that such
two-phase accretion flows are more strongly rotating, or more
precisely, that their circularisation radii are larger (see
Section~\ref{sec:circ}).

\begin{figure}
\centerline{\psfig{file=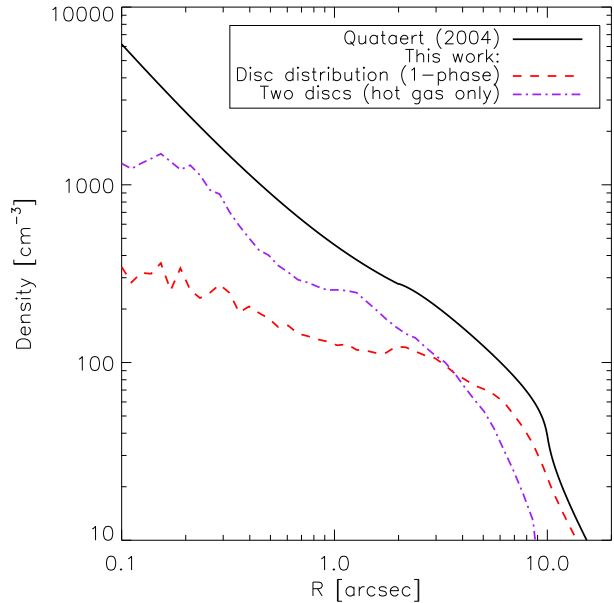,width=.49\textwidth,angle=90}}
\caption{Hot gas density profiles of `one-phase' simulations with
stars arranged in disc orbits (red dashed line,
Section~\ref{sec:moving}) and two-phase simulations with two discs of
stars (violet dot-dashed, Section~\ref{sec:2discs}). Note that the
two-phase simulation yields a much more centrally concentrated
profile.}
\label{fig:nden_lbvwr}
\end{figure}

The other notable difference between the one and two phase simulations
is that the latter produce much more concentrated radial density
profiles for hot gas (Fig.~\ref{fig:nden_lbvwr}). Due to the
presence of slow winds and thus more prominent radiative cooling, the
two-phase simulations yield less pressure support for the gas. A
larger angular momentum is then needed to support the gas against
gravity, as pointed out above. However this extra support is
apparently insufficient and the radial density profile plunges in
quicker in the two-phase case than it does in the one phase
case. Interestingly, despite using a different setup and including
different physics, our two-phase hot gas density curve is by and large
similar in shape to that of \cite{Quataert04}, although yielding a
lower accretion rate.

\section{What would Chandra see?}\label{sec:chandra}

Having obtained maps of the gas density and temperature distributions,
we calculate the expected X-ray emission of each pixel. We use an
optically thin cooling function $\Lambda(T)$ (see Section~\ref{sec:method}), modified to crudely
model the effects of the interstellar absorption. As is well known,
the column depth of the cold ISM to \sgra\ is very high, $N_{\rm H}
\sim 10^{23}$ hydrogen atoms cm$^{-2}$ \citep[e.g.,][]{Baganoff03a}, and
hence {\em Chandra} receives hardly any photons with energy less than $\sim
2$ keV. Therefore, gas cooler than $T\sim 10^7$ Kelvin will contribute
very little to the X-ray counts. We then use
\begin{equation}
\Lambda_{\rm mod}(T) = \frac{\Lambda(T)}{1 + (T_{\rm cut}/T)^5}\;,
\label{lambdam}
\end{equation}
where $T_{\rm cut}=10^7$ Kelvin. \footnote{A very sharp rollover in
temperature is justified by the very strong dependence of
photo-electric absorption cross section on photon energy.} Below we
consider X-ray emission from two of our simulations.

\begin{figure}
\centerline{\psfig{file=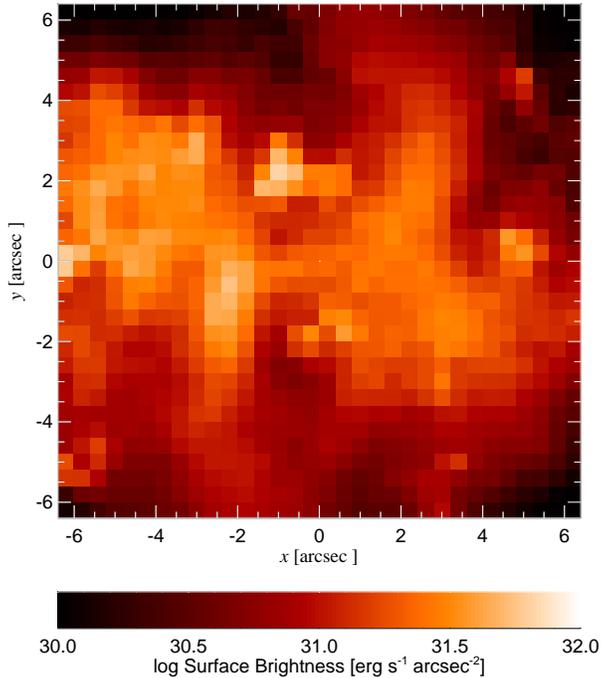,width=.49\textwidth,angle=0}}
\caption{Simulated {\em Chandra} view of the inner $12''$ by $12''$ for
the simulation with fast winds only (Section~\ref{sec:fast}), with 40
stars arranged in a disc configuration. The corresponding density
profile was shown in Fig.~\ref{fig:ndenrot} (dashed red curve). Note
the strong and highly extended emission from stellar wind shocks. No
central source associated with \sgra\ is actually visible.}
\label{fig:chandra1}
\end{figure}

Figure \ref{fig:chandra1} shows a resulting X-ray map for the
simulation with fast winds only (Section~\ref{sec:fast}), with 40
stars in a rotating disc configuration.  The map shows the luminosity
(normalised to one arcsec) integrated over pixels of $0.5''$, which is
about the {\em Chandra} pixel size (clearly we could have produced much
finer simulated X-ray images of the GC). Remarkably, the density of
the hot gas in the inner arcsecond is so low that stellar winds
actually dominate the X-ray map. This is due to two factors. First,
the stellar density is greater for stars located in a (relatively
thin) disc than it is for the same stars spread around a spherical
shell at the same radial distance.  As a result, the stars are closer
to each other and the shocked gas density is higher. Second, the large
angular momentum of the gas prevents most of it from flowing into the
inner arcsecond of this simulation (see Figs.  \ref{fig:angmom} and
\ref{fig:ndenrot}), thus reducing the X-ray emission from the centre.

\begin{figure}
\centerline{\psfig{file=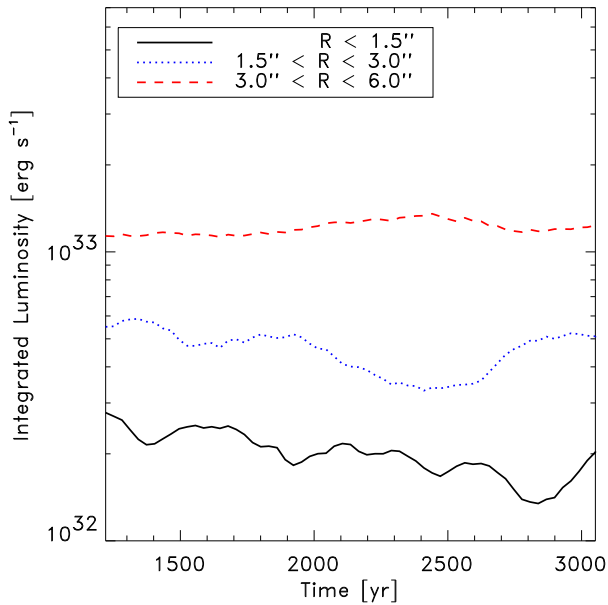,width=.49\textwidth,angle=90}}
\caption{Light curves of three radially selected regions, as
indicated in the inset. Variability is due to changes in the relative
positions of the stars. The luminosity of the inner region is
correlated with the accretion rate, because both are determined by the
density in the inner arcsecond.}
\label{fig:chandra1a}
\end{figure}

Figure \ref{fig:chandra1a} depicts the light curves of the radial
annuli ($R < 1.5$, $3$ and $6''$, respectively) for the simulation
corresponding to Fig. \ref{fig:chandra1}. Clearly, the central
region never dominates the X-ray emission. Also, variability is rather
mild and occurs on time scales of hundreds of years.

\begin{figure}
\centerline{\psfig{file=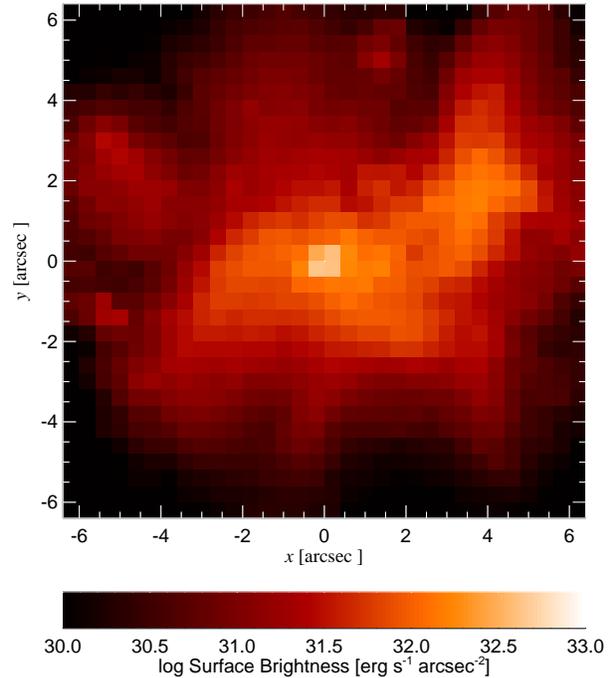,width=.49\textwidth,angle=0}}
\caption{Simulated {\em Chandra} view of the inner $12''$ by $12''$ for
the simulation shown in Fig. \ref{fig:largeview}. The rings of stars
are inclined at a $45^\circ$ angle to the line of sight in this
figure. Note that the central arcsecond clearly stands out in X-rays,
but the emission is spatially extended, perhaps slightly more than in
the real \sgra\ observations by {\em Chandra}.}
\label{fig:chandra2}
\end{figure}

Figure \ref{fig:chandra2} shows the more `realistic' simulation which
includes both fast and slow winds with stars located in two rather
than one ring. Note that the scale of the colour bar on the bottom is
10 times higher than that used in Fig.~\ref{fig:chandra1}. In sharp
contrast with Fig. \ref{fig:chandra1}, the central source clearly
stands out now at a level consistent with that observed by
{\em Chandra}. This difference is caused by a much stronger concentration of
the hot gas in the centre in this latter model, discussed in
Section~\ref{sec:1vs2}.

\begin{figure}
\centerline{\psfig{file=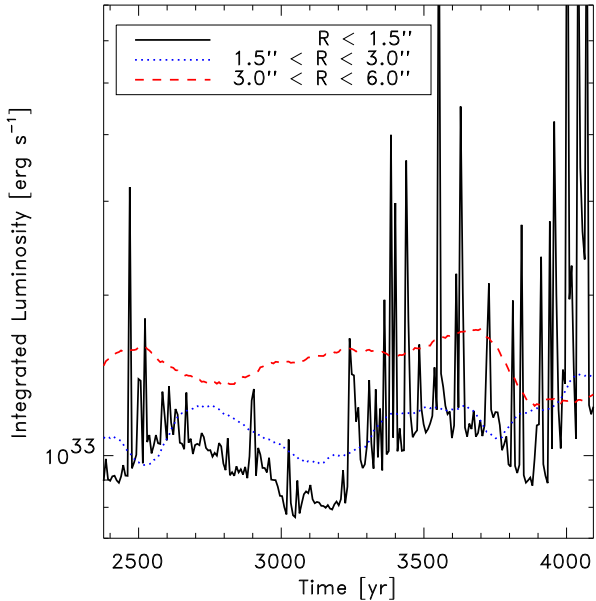,width=.49\textwidth,angle=90}}
\caption{The same as for Fig.~\ref{fig:chandra1a} but for the
simulation with two rings of stars (see Fig.~\ref{fig:chandra2}). Note
that now the central region is brighter and much more variable. The
variability in the light received from the central region is mainly caused
by cool blobs of gas raining down on to the disc, shock-heating gas to
X-ray emitting temperatures.}
\label{fig:chandra2a}
\end{figure}

Figure~\ref{fig:chandra2a}, shows the light curves corresponding to
Fig. \ref{fig:chandra2} and can be directly compared with
Fig.~\ref{fig:chandra1a}. We note that in the more `realistic'
simulation the X-ray luminosity of the central source is higher and is
by far more variable. The steady state X-ray emission is larger simply
due to the higher hot gas density in the inner region, whereas the
variability is produced by shock-heated blobs impacting the cold
disc. During such events the peak in the X-ray intensity can actually
shift from \sgra\ nominal position by $\sim 0.5''$ or so.

It would be interesting to compute the X-ray spectrum of each pixel,
and then compare that to {\em Chandra} observations of \sgra. This would
involve calculation of the local radiation spectrum, radiation
transfer along rays and a model for X-ray absorption in the
interstellar matter between the GC and us. We leave such a detailed
comparison with observations for future work.

\section{Discussion} \label{sec:discussion}

The theory of accretion flows on to SMBHs is an area of active
research \citep[e.g.,][]{Narayan02} where observational tests
constitute strong drivers in the field. Due to its proximity, \sgra\
is the only SMBH where it is becoming possible to constrain
observationally the accretion flow properties both at small and large
radii simultaneously \citep[$R\sim 10-100R_{\rm S} $ and $R\sim 10^5
R_{\rm S}$; see][]{Bower03,Baganoff03a,Nayakshin05}.

The most direct method to measure the accretion rate is through {\em
Chandra} X-ray observations \citep{Baganoff03a}. However, this method
has the following drawbacks: (a) gas cooler than $\sim 10^7\,{\rm K}$
cannot be observed due to a very high neutral hydrogen absorbing
column to the GC, thus the contribution of cool gas is unknown; (b)
bulk motions of the gas can in principle make some of the gas unbound,
but these do not directly enhance the X-ray emission, and thus they
can be missed; (c) only the instantaneous conditions can be probed,
whereas the relevant time scales can be tens and hundreds of years.

The second method to infer the input accretion rate is through
observations of stellar winds and models of the outflows and gas
accretion on to the SMBH.  This method alleviates deficiencies
(a-c) discussed above, but of course comes with its own set of
problems and uncertainties. In this paper, we have attempted to reduce
these in the part of the theoretical (numerical) modelling of the wind
hydrodynamical evolution. Conceptually, we built on the previous
numerical work in the area \citep{Coker97,Rockefeller04} and on the
semi-analytic model of \cite{Quataert04}. However, we have used a
Lagrangian SPH/$N$-body code, allowing us to follow the problem in its
full 3D setting. Compared with previous authors, we have been able to
relax assumptions that stellar wind sources are fixed in space and
that shocked winds suffer no radiative losses. We have also varied
stellar orbital distributions, testing rotating disc-like and
isotropic distributions.

\subsection{Reduction of accretion due to anisotropy and net angular momentum} 

We performed three runs with fast stellar winds ($v_{\rm w} = 10^3$ km
s$^{-1}$) produced by 40 stars. These runs tested the importance of
the orbital motions and the source distribution. For a stationary
spherical stellar distribution, the accretion rate (see Fig.
\ref{fig:accraterot}) was found to be around $\sim 10^{-5}
\msun$~year$^{-1}$, with variability at a level of 10\%. These results
are consistent with that of \cite{Coker97} after correcting for small
differences in wind velocities and stellar wind loss rates. Allowing
the stars to follow randomly oriented circular Keplerian orbits
decreased the accretion rate by a factor of three, and increased the
variability up to $\sim 50-70$\% (blue dotted curve in
Fig.~\ref{fig:accraterot}).  Placing these same stars into a disc
reduced the accretion rate by another factor of three or so, while
keeping the variability magnitude at about the same level. The main
driver of the differences in these three tests is the net angular
momentum of the gas (Fig. \ref{fig:angmom}).  We find that only the
gas with a low enough angular momentum makes it to the inner boundary
and hence is accreted, and the fraction of such gas decreases when the
source distribution is rotating with a common direction (a disc).

Another way to view this result is to say that the angular momentum of
the gas and `random' gas motions reduce the gas capture
radius. Although authors' definition of the latter differ slightly,
most consider the capture radius to be 1--$2''$
\citep{Baganoff03a,Rockefeller04}, whereas we would define $R_{\rm
capt}$ to be about $0.4''$ based on our simulations. Indeed the gas can
be called captured only at this region, where accretion (see
Fig. \ref{fig:massflow}) strongly dominates; at large radii outflow
and inflow nearly cancel each other.

\subsection{Cool phase of stellar winds}

In a fixed pressure environment, gas cooling time is a very strong
function of temperature: $t_{\rm cool} \propto T/(\Lambda(T) n)
\propto T^2 /\Lambda(T)$, where $\Lambda(T)$ is the optically thin
cooling function, here dominated by metal line cooling. Further,
shocked gas temperature is proportional to the initial velocity of the
winds squared, $v_w^2$. Taking these facts together one finds that gas
cooling time scales as $v_w^{5.4}$ \citep[eq. 2]{C05}.  Therefore, for
the Wolf-Rayet winds, with velocities of a thousand km s$^{-1}$, gas
cooling time is much longer than the dynamical time and the effect of
cooling is negligible. However, for outflow velocity of 300 km
s$^{-1}$, radiative cooling is faster than the local dynamical
time. We therefore find that these winds cool radiatively and form
clumps of cold gas.

We find that cold clumps form independently of the geometry of the
stellar system, as long as wind velocity is low enough. The morphology
of cold gas however strongly depends on the orbital distribution of
slow wind sources. In Section~\ref{sec:twophase} we tested two extreme
cases for the stellar distribution: stars in two discs, and in a
spherically symmetric system. An extended cold disk was present in the
former configuration, and only a tiny gas ring in the former.

A detailed comparison between the gas morphology in our simulations
and in the observations is beyond the scope of this work. However, it
is interesting to note that the radio images presented by
\cite{Wardle92} show an asymmetry in the gas that these authors
attributed to the influence of stellar winds. This anisotropy suggests
that the stellar wind sources are not distributed isotropically,
whereas the absence of a conspicuous disc points out that the narrow
line stars of the inner few arcsec have different orbital planes.

Eckart \& Morris have recently reported\footnote{KITP Conference on
the GC, talks available online at
http://online.kitp.ucsb.edu/online/galactic\textunderscore c05 .} the
presence of extended mid-infrared emission blobs. Their red spectrum
is interpreted as dust emission. It is possible that some of these
blobs were actually formed from the stellar winds as in our
simulations. Since for the slow winds in our simulations cooling is
important, and since the gas would probably cool much further than the
minimum temperature of $10^4$ K assumed here, formation of dust would
follow. One of the observed blobs is only $0.026''$ away from \sgra\
in projection and is probably responsible for the offset between the
dynamical centre and the \sgra\ mid-infrared emission detected by
\cite{Clenet04}. Monitoring its proper motion will show whether it is
physically close to \sgra\ and therefore if it could be identified
with one of the clumps that feed the SMBH in our simulations.

{  Finally, radio observations \citep{Yusef98} have revealed the
presence of clumps with proper motions high enough to escape from the
GC region. The mass of one of them, `the bullet', is estimated at
$8\times10^{-4} \msun$. This mass is of same order as the typical mass
of cold clumps leaving the inner region in our simulations.}

\subsection{Long term evolution of the disc}

In Section~\ref{sec:2discs} we stopped the simulation after $\sim$
4000 yr. At this point the total mass of cold gas keeps growing in the
inner few arc-seconds and there is no obvious tendency for reaching a
steady state. However, because of factors not included in the
simulation, it is very likely that in a few thousand years the disc
would be destroyed. One reason for that is the presence of many
massive stars that would explode as a supernova (SN) within a short
time-scale. There are currently tens of WR stars in the GC star
cluster. This stellar phase is expected to last a few $\times 10^5$
yr, at the end of which the star explodes as a SN. Therefore a SN
explosion is expected every $\sim 10^4$ yr.

Let $R_{\rm sn}$ be the distance between a supernova and \sgra, and $E
= 10^{51} E_{51}$ erg s$^{-1}$ be the total energy content of the supernova
shell. The disc material will then be accelerated to a velocity
$v_{\rm acc}$ given by $\Sigma_{\rm disc} v_{\rm acc}^2 \sim E_{\rm
sn}/4\pi R_{\rm sn}^2$. For example, for disc radius $R_{\rm disc} =
2''$, the ratio of this velocity to the local Keplerian velocity is
\begin{equation}
\frac{v_{\rm acc}^2}{v_{\rm K}^2} \sim 1 \;\frac{E_{51}}{M_{10} R_{10}^2}\;,
\label{vacc}
\end{equation}
where $R_{10}= R_{\rm sn}/10''$ and $M_{10} = \pi \Sigma_{\rm disc}
R_{\rm disc}^2$ in units of 10~$\msun$. This shows that a supernova
occurring within the inner $0.5\,$pc of the Galaxy would destroy the
disc. A fraction of the disc with the lowest angular momentum would
likely be captured by the SMBH and would result in a bright flare that
would last at least a few hundred years, the dynamical time at a couple
of arc-seconds distance from \sgra. Such flares may be responsible for
the X-ray/$\gamma$-ray echo of \sgra\ recent activity on nearby giant
molecular clouds, most notably Sgr B2 \citep{Sunyaev93, Koyama96,
Revnivtsev04}, and the observed plumes of hot gas also indicative of
former AGN activity \citep{Baganoff03a}.

Another factor that should be taken into account for a realistic
simulation on longer time-scales is the presence of the
`mini-spiral'. This seems to be a collection of infalling gas
structures, with masses of dozens of $\msun$ \citep{Paumard04}. Since
their orbital time-scales are a few $\times 10^4$ years, by that time
collisions between these structures and the disc are expected,
resulting in the destruction or a significant re-arrangement of the
latter. Again this would enhance the accretion rate on to \sgra.

\subsection{Variability}

We have shown that adding more realistic ingredients in our
simulations introduces substantial variability into the accretion rate
on to \sgra. In Section~\ref{sec:moving}, with 40 identical stars, the
variability increased from less than $\sim10\%$ to about $50\%$ by
just allowing the stars to follow circular orbits
(Fig.~\ref{fig:accraterot}). Further, in Section~\ref{sec:2discs}, the
inclusion of slow stellar winds led to the formation of cold
clumps. The accretion of these clumps proceeds in short bursts,
producing a highly variable transfer of mass into the inner $0.1''$
(Fig. \ref{fig:accratelbvwr}).  The actual rate of accretion on to the
SMBH has still to be determined by the accretion flow physics that we
cannot resolve here, but it is expected to maintain most of the
variability in time-scales longer than $\sim 100$ yr.

There are additional sources of variability that have not been
included in our treatment. Some of the stars with high outflow rates
probably have orbits with a non-negligible eccentricity. Then the
fraction of winds that can be captured from a given star by the black
hole changes with time. In addition, there could be intrinsic
variability of the stars themselves. LBVs outside the GC have been
observed to vary their mass-loss rates by more than an order of
magnitude within a few years \citep[e.g.,][]{Leitherer97}. In the GC,
the line profiles of IRS 16 NE \& NW have changed in the last few
years. This is usually attributed to orbital acceleration, but it
could be partially caused by changes in the winds properties (Najarro,
priv. comm.). {  Variability can also be produced by close X-ray
binaries. Recently, \cite{Muno05} and \cite{Porquet05} identified a
transient source located within 0.1 pc of the GC as a low-mass X-ray
binary. \cite{Bower05} detected its radio counterpart, and argue that
it is the signature of a jet interacting with the dense gas of Sgr~A
West. Such a jet can influence the kinematics of the gas, and in some
cases could drive material to the black hole vicinity.} Finally, as we
noticed in the previous subsection, SNe or the infall of cold gas from
larger scales could lead to a strong increase in the accretion rate.

In the extreme sub-Eddington regime of \sgra\ accretion, the
dependence of the luminosity on the accretion rate is very non-linear
\citep[see e.g.,][]{Yuan02}. Thus, even small changes in the accretion
rate could result in a strongly enhanced X-ray emission.  The results
from our simulations, the observational evidence for higher luminosity
in the recent past \citep{Sunyaev93, Koyama96, Revnivtsev04}, and the
idea of star formation in an AGN-like accretion disc a few million yr
ago \citep[e.g.,][]{Levin03,NC05}, all suggest that on long
time-scales \sgra\ is an important energy source for the inner Galaxy.

\subsection{Circularisation of the flow}\label{sec:circ}

The degree to which accretion flows are rotating is very important for
theoretical models of these flows. For example, \cite{Melia92,Melia94}
assumes that gas inflows in essentially a quasi-spherical
\citep{Bondi52} manner down to a circularisation radius of $R_{\rm c} \sim
100 R_{\rm S}$, whereas, e.g., \cite{Narayan02} and \cite{Yuan03} assume that the flow
is strongly rotating already at a sub-arcsecond region.  Recent
detailed hydro and MHD simulations of \cite{Proga03a,Proga03b}
confirmed that the nature of the accretion flow strongly depends on
the angular momentum of the gas at the outer boundary of the flow.

Here we find that in all of our simulations the accretion flows posses
a relatively large angular momentum, and that in the two-phase wind
simulations it is larger than in the one phase case. In terms of
circularisation radii for the flows, we find $R_{\rm c}\sim 0.01'' \sim 10^3
R_{\rm S}$ for one-phase winds, and $R_{\rm c}\sim 0.1''\sim 10^4 R_{\rm S}$ for
two-phase flows. Physically, one-phase flows are hotter, and any
significant rotation unbinds the gas in the inner arcsecond. This
works as a surprisingly effective selection criterion for determining
which particles do accrete, so in our one-phase simulations the
gas in the inner region has roughly the same angular momentum,
independent of the stellar motion.

On the other hand, in the two-phase simulations (fast and slow winds),
radiative cooling of denser regions reduces the mean temperature of
the gas. As a result, less pressure support is available for the gas,
and the `selection' criterion does not operate anymore. These flows
rotate at almost the local circular Keplerian velocity near the inner
boundary of our simulations. Remarkably, this result (for the hot gas)
is largely independent of the orbital distribution of the stellar wind
sources. Therefore, unless the mass loss rates of the slow (narrow
line) wind stars are grossly over-estimated observationally, we
conclude that weakly rotating accretion flows are unlikely to form in
\sgra.

\section{Conclusions}\label{sec:conclusions}

In this paper we presented a detailed discussion of our new numerical
simulations of wind accretion on to \sgra. Compared with previous
works, our methodology includes a treatment of stellar orbital motions
and of optically thin radiative cooling.  While the results are
strongly dependent on the assumptions about stellar mass loss rates,
orbits, and wind velocities, some relatively robust conclusions can be
made.

Unless mass loss rates of narrow line mass losing stars
\citep{Paumard01} are strongly over-estimated, the gas at $r\sim 1''$
distances from \sgra\ has a two-phase structure, with cold filaments
immersed into hot X-ray emitting gas. Depending on the geometry and
orbital distributions in the mass-loosing star cluster, the cold gas
may be settling into a coherent structure such as a disc, or be torn
apart and heated to X-ray emitting temperatures in collisions. Both
the fast and the slow phase of the winds contribute to the accretion
flow on to \sgra. The accretion flow is rotating rather than
free-falling in sub-arcsecond regions, with a circularisation radius
of order $10^4 R_{\rm S}$. The accretion rates we obtain are of the
order of $3\times 10^{-6} \msun$~year$^{-1}$, in accord with the {\em
Chandra} observations \citep{Baganoff03a}. The accretion of cooler gas
proceeds separately via clump infall and is highly intermittent,
although the average accretion rate is dominated by the quasi-constant
inflow of hot gas. As is true for the hot gas, most of the cold gas
outflows from the simulated region. However some of this gas is bound
to \sgra, and, had we resolved a larger region, would come back into
the inner regions on eccentric orbits.

A generic result of our simulations is the large variability in the
accretion rate on to \sgra\ on time scales of the order of the stellar
wind sources orbital times (hundreds to thousands years). Even in the
case of one-phase flows the accretion rate shows variability by
factors of a few. Simulations including more diverse populations of
winds and stellar orbits, the presence of the single, very important
mass-loosing star cluster IRS13 \citep{Maillard04, Schoedel05}, or the
observed ionised (e.g. $T\sim 10^4$ K) gas, can all be expected to
further increase the time-dependent effects. This implies that the
current very low luminosity state of \sgra\ may be the result of a
relatively unusual quiescent state. It also means that the real
time-averaged output of \sgra\ in terms of radiation and mechanical
jet power may be orders of magnitude higher than what is currently
observed. The role of \sgra\ for the energy balance of the inner
region of the Galaxy may therefore be far more important than its
current meager energy output would suggest.  Observations of
$\gamma$-ray/X-ray echos of past activity of \sgra\
\cite[e.g.,][]{Revnivtsev04} seem to confirm these suggestions from
our simulations.

Further improvements in the models of wind accretion on to \sgra\ will
benefit most strongly from better observational constraints on the
properties of the stellar winds in the central parsec of the Galaxy.
Future numerical improvements may be a higher dynamical range between
$R_{\rm in}$ and $R_{\rm out}$, a larger number of SPH particles, and
inclusion of magnetic fields. The latter are now believed to be very
important for gas angular momentum transport via generation of the
magneto-rotational instability \citep[e.g.,][]{Balbus91}. At the same
time, we expect angular momentum transport to be important only inside
or close to the circularisation radius, i.e. somewhere near our inner
boundary.

\section*{Acknowledgements}

JC acknowledges D.~{\v S}ija{\v c}ki and C.~Scannapieco's help during
his first attempts to use {\small GADGET-2}. We thank E.~Quataert for
sending us data files from his calculations and the anonymous referee
for useful suggestions. Finally, this work benefited from discussions
with the members of the GC group at MPE, especially R.~Genzel and
T.~Paumard, and with F.~Baganoff, M.~Morris, F.~Najarro and
F.~Yusef-Zadeh.

\bibliography{biblio}
\bibliographystyle{mnras}

\end{document}